\documentclass[prx,amsmath,amssymb,twocolumn]{revtex4-2}
\usepackage{graphicx}
\usepackage{subfigure}
\usepackage{adjustbox}
\usepackage{bm}
\usepackage{color}
\usepackage{braket}
\usepackage{standalone}
\usepackage{multirow}
\usepackage{tikz}
\usepackage{mathrsfs}
\usepackage{dsfont}
\usepackage{comment}
\usepackage{amsmath}

\usepackage[colorlinks,bookmarks=true,citecolor=blue,linkcolor=red,urlcolor=blue]{hyperref}

\newcommand{\be}{\begin{equation}}
\newcommand{\ee}{\end{equation}}

\newcommand{\ba}{\begin{eqnarray}}
\newcommand{\ea}{\end{eqnarray}}

\newcommand*{\id}{{\normalfont\hbox{1\kern-0.15em \vrule width .8pt depth-.5pt}}}
\newcommand{\p}{\partial}
\newcommand{\la}{\label}
\newcommand{\Tr}{\text{Tr}}

\begin{document}

\title{Very high-energy collective states of partons in fractional quantum Hall liquids}

\author{Ajit C. Balram$^{1,2}$, Zhao Liu$^{3}$, Andrey Gromov$^{4}$, and Zlatko Papi\'c$^{5}$}
\affiliation{$^{1}$Institute of Mathematical Sciences, CIT Campus, Chennai 600113, India}
\affiliation{$^{2}$Homi Bhabha National Institute, Training School Complex, Anushaktinagar, Mumbai, 400094, India}
\affiliation{$^{3}$Zhejiang Institute of Modern Physics, Zhejiang University, Hangzhou 310027, China}
\affiliation{$^{4}$Brown Theoretical Physics Center and Department of Physics, Brown University, Providence, Rhode Island 02912, USA}
\affiliation{$^{5}$School of Physics and Astronomy, University of Leeds, Leeds LS2 9JT, United Kingdom}

\date{\today}

\begin{abstract}
The low-energy physics of fractional quantum Hall (FQH) states -- a paradigm of strongly correlated topological phases of matter -- to a large extent is captured by weakly interacting quasiparticles known as composite fermions. In this paper, based on numerical simulations and effective field theory, we argue that some \emph{high-energy} states in the FQH spectra necessitate a different description based on \emph{parton} quasiparticles. We show that Jain states at filling factor $\nu{=}n/(2pn{\pm}1)$ with integers $n,p{\geq}2$ support two kinds of collective modes: In addition to the well-known Girvin-MacDonald-Platzman (GMP) mode, they host a high-energy collective mode, which we interpret as the GMP mode of partons. We elucidate observable signatures of the parton mode in the dynamics following a geometric quench. We construct a microscopic wave function for the parton mode, and demonstrate agreement between its variational energy and exact diagonalization. Using the parton construction, we derive a field theory of the Jain states and show that the previously proposed effective theories follow from our approach. Our results point to partons being ``real'' quasiparticles which, in a way reminiscent of quarks, become observable only at sufficiently high energies.

\end{abstract}

\maketitle

\section{Introduction} 

Since their discovery, fractional quantum Hall (FQH) phases continue to attract attention for their exotic topological properties~\cite{Tsui82, Laughlin83, Halperin82, WenEdge, Moore91}, including the recent experimental observations of fractional statistics of their underlying charged quasiparticles~\cite{Nakamura2020, Bartolomei173}. A highly successful theoretical approach for a large class of FQH states is based on reformulating the problem in terms of weakly interacting quasiparticles known as composite fermions (CFs)~\cite{Jain89}. The CF approach exists in its microscopic and field-theoretic incarnations. The microscopic formulation has proven to be extremely successful in producing accurate wave functions for the ground state and low-lying excited states~\cite{Jain07}. The field theory due to Halperin, Lee, and Read (HLR)~\cite{halperin1993theory} predicts the composite Fermi liquid (CFL) nature of the gapless state in the half-filled Landau level (LL) and describes the underlying mechanisms for the formation of non-Abelian paired states~\cite{Moore91, Read00}. 

Recently, the field theory of CFs underwent a major transformation when Son~\cite{Son15} explained how the lowest LL (LLL) projection could be incorporated into the HLR theory. This was accomplished by noting that particle-hole (PH) symmetry becomes an exact symmetry of the FQH problem if the electrons are restricted to the LLL. The PH symmetry is implemented utilizing a Dirac description of CFs and a massless version of the Galilean invariance~\cite{Golkar16}. Among the successes of the Son-Dirac theory~\cite{Son15} is the calculation of the spectrum of collective modes~\cite{golkar2016higher} near filling factor $\nu{=}1/2$ and the calculation of the projected static structure factor~\cite{nguyen2018fractional}, in accordance with the general Ward identities~\cite{Nguyen17, Gromov14}. The Son-Dirac theory~\cite{Son15} applies to FQH states in the vicinity of $\nu{=}1/2$. More generally, for FQH states in the Jain sequence of filling factors $\nu {=}n/(2pn{\pm}1)$, this approach is not valid due to the lack of PH symmetry: States around $\nu{=}1/(2p)$ get mapped to states near $\nu{=}(2p{-}1)/(2p)$ upon PH transformation. Several kinds of effective theories are proposed to treat the general Jain states with $p{>}1$~\cite{wang2016composite, you2018partially, Goldman18, wang2019dirac, Kumar19}.

One crucial piece of data needed to understand the general Jain sequence is encoded in the neutral collective modes they support. Since the work of Girvin, MacDonald, and Platzman (GMP)~\cite{Girvin85, Girvin86}, it has been known that all FQH fluids support a collective excitation that can be viewed as a (LLL-projected) density wave on top of the electronic ground state -- the gapped analog of the roton excitation in liquid helium~\cite{FeynmanStatistical}. This collective mode has a microscopic description in terms of a ``single-mode approximation" (SMA), which has been extensively tested numerically~\cite{Renn93, Yang12b, Repellin2014} and detected experimentally using inelastic light scattering~\cite{Pinczuk93, Platzman96, Kang01, Kukushkin09}. Recent work~\cite{Haldane11, Gromov17, nguyen2018fractional} points out that the long-wavelength limit of the GMP mode exhibits a novel property: As the momentum $q{\to}0$, the density wave excitation can be interpreted as an emergent quantum geometry. This geometric degree of freedom is dubbed the ``FQH graviton," since it carries angular momentum $L{=}2$~\cite{Yang12b, Golkar2016, Gromov17, gromov2017investigating} suggestive of the spin-$2$ elementary particle. The FQH graviton is hidden from conventional probes for collective modes such as inelastic light scattering~\cite{Pinczuk93, Platzman96, Kang01, Kukushkin09}, since these probes are limited to finite momenta. In contrast, recent works~\cite{Liu18, lapa2019geometric, Liu2021, Kirmani2021} show that the graviton can be directly excited in a dynamical quench experiment, where the band mass or dielectric tensor of the two-dimensional electron gas is suddenly made anisotropic or the magnetic field is abruptly tilted (see also a recent proposal using surface acoustic waves~\cite{KunYang2016}). 

An important clue about how to approach the general Jain series comes from a recent work \cite{Nguyen2021}, where it is demonstrated that the theories of Refs.~\cite{Goldman18, wang2019dirac}, as written, are inconsistent with LLL projection, because they violate the bound on the projected static structure factor discovered by Haldane \cite{HaldaneStructureFactor}. The authors of Ref.~\cite{Nguyen2021} argue that the only way to remedy the field theory is to \emph{postulate} an existence of a collective spin-$2$ mode similar to the usual GMP magnetoroton mode. This mode supplies the missing contribution to the structure factor and ensures the consistency of the field theory with LLL projection.

In this paper, we study the neutral collective modes of the general Jain series at filling $\nu{=} n/(2pn \pm 1)$ using the \emph{parton} construction, at both the microscopic and field-theoretic levels. The parton picture of elementary particles was originally introduced by Feynman to describe the high-energy physics of hadrons~\cite{FeynmanPartons,kogut1973parton}. In the FQH context, the microscopic version of the parton construction was introduced by Jain~\cite{Jain89b} as a generalization of the CF theory to describe a wider class of FQH states, especially even denominator states, while its field-theoretic version was formulated by Wen~\cite{Wen91} to provide a mean-field derivation of certain non-Abelian FQH states. We describe a new parton construction that naturally incorporates the LLL projection. In our parton theory, the LLL \emph{Dirac} electron is broken down into two partons $\Psi_e {=} \Psi_p \cdot \phi$, where the fermionic parton $\Psi_p$ forms a Jain state in the primary sequence at filling $\nu {=} n/(2n \pm 1)$, while the bosonic parton $\phi$ forms a Laughlin state at filling $\nu {=}1/[2(p-1)]$. This construction implies the effective theories studied in Refs.~\cite{Goldman18, wang2019dirac, Nguyen2021}.

\begin{figure}[bht]
  \centering
  \includegraphics[width=\linewidth]{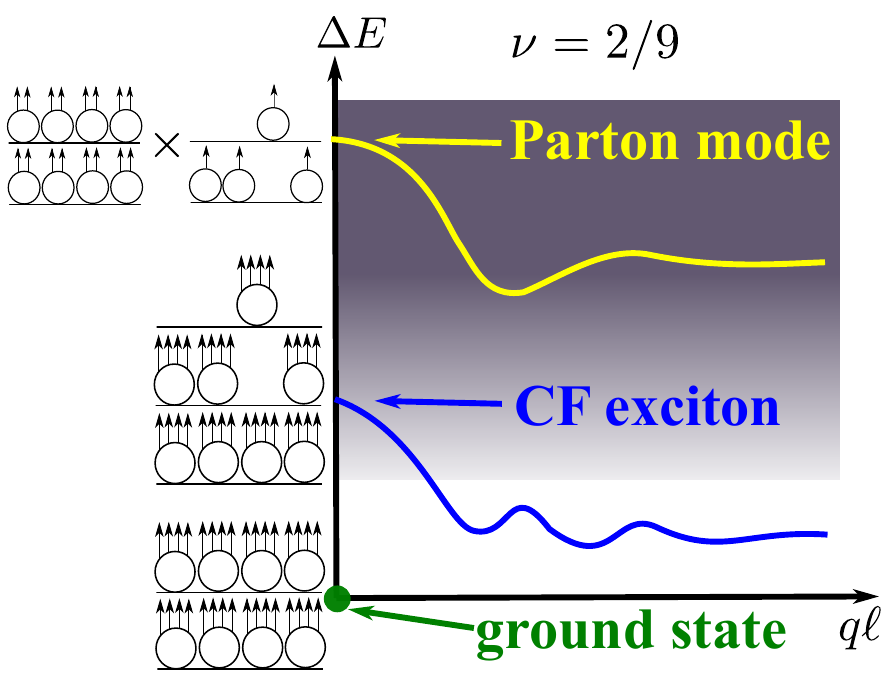}
  \caption{Schematic representation of the spin-$2$ collective modes of the $\nu{=}2/9$ Jain state. The low-energy CF exciton mode is shown in blue, while the high-energy parton mode is shown in yellow. The green dot represents the ground state, and the shaded region represents the continuum of excitations. The circle and arrows together denote a composite fermion [boson] which is a bound state of an electron (circle) and an even [odd] number of vortices (arrows). In the long-wavelength limit, we also refer to the CF exciton mode as the GMP mode, as the two describe similar physics in this limit. }
  \label{fig: schematic_modes}
\end{figure}

The parton construction naturally leads to two neutral collective modes: the GMP mode of the fermionic parton and the GMP mode of the bosonic parton; see Fig.~\ref{fig: schematic_modes} for an illustration. The former mode is also captured by the CF theory in the form of a CF exciton -- promoting one CF from the highest filled CF-Landau level to the lowest unoccupied CF-Landau level~\cite{Jain07}. In the long-wavelength limit of our interest, the two approaches, GMP and CF exciton, yield similar results~\cite{Yang12b, Repellin2014}, and we use the two terms interchangeably. In contrast, the focus of the present paper is the GMP mode of bosonic partons, a new type of excitation that is not captured by the GMP and CF theories.

The paper is organized as follows. In Sec.~\ref{sec: EFT} we discuss the parton field theory, rederive the previous results of Refs.~\cite{Goldman18, wang2019dirac, Nguyen2021}, and explain the origin of the collective modes, paying special attention to the LLL projection. In Sec.~\ref{sec: parton_wf} we give the microscopic parton construction and propose LLL-projected trial wave functions for the collective modes, evaluating their variational energies. In Sec.~\ref{sec: dynamical_sf} and \ref{sec: quench_dynamics}, we perform a detailed numerical study of the collective modes at $\nu{=}2/7$ and $2/9$. We show that the two modes can be distinguished not only by their energy, but also by their chirality [in the case of states in the Jain sequence $n/(2pn-1)$] as well as by their clustering properties. We also study the geometric quench and observe two independent oscillation frequencies that indicate the existence of two modes. Similarly, we show that the dynamical structure factor, at low wave numbers, is heavily concentrated on the two collective modes, despite the bosonic parton mode occurring at very high-energy. Our results provide evidence that partons emerge as fundamental quasiparticles of the $p{>}1$ Jain FQH states when probed at sufficiently high energies. Moreover, these results highlight the qualitatively different physics between states lying in the primary and secondary Jain sequences. We conclude in Sec.~\ref{sec: discussion_conclusions} by discussing some open problems and future directions. Appendixes contain further derivations and numerical data for other filling factors, including FQH states of bosons. 

\section{Effective field theory}
\label{sec: EFT}
One approach to an effective field theory for general Jain FQH states is based on the requirement that in the parent CFL state the CF has a fractional Berry phase around its Fermi surface~\cite{wang2016composite, you2018partially}. A different approach postulates that the Dirac CF with Berry phase $\pi$ interacts with a Chern-Simons (CS) term with the coefficient $(p-1)/(8\pi p)$ \cite{Goldman18, wang2019dirac}. The term is motivated partly by the empirically observed ``reflection symmetry'' of the $I-V$ curves near $\nu {=}1/(2p)$~\cite{shahar1995universal, Shimshoni97}. The CS term manifestly breaks PH symmetry; however, it is not clear what forces the CF to have a Berry phase of $\pi$. Still, the $- 2\pi \nu$ Berry phase at filling $\nu{=}1/4$ is demonstrated numerically \cite{wang2019dirac}. The study of the Berry curvature shows that the $\pi$ Berry phase comes from a singularity at momentum $0$, while ${-}\pi(p{-}1)/p$ is smoothly distributed in momentum space. 

In this section, we expose a field-theoretic version of the parton construction. This will allows us to derive the effective Lagrangian for the Jain series at filling $\nu {=} n/(2pn{\pm} 1)$ from a microscopic starting point and an appropriate mean-field approximation. The two main ingredients in the construction are (i) parton decomposition of the electron operator and (ii) LLL projection implemented via the duality discovered by Son~\cite{Son15}. The effective theories studied in Refs.~\cite{Goldman18, wang2019dirac} follow from our construction. We further show that the extra spin-$2$ mode discussed in Ref.~\cite{Nguyen2021} corresponds to the collective mode of partons. We assume that our theories are defined on $\mathbb R^3$ and neglect non-local information. Careful discussion of the parton construction can be found in Refs.~\cite{1999-Wen, 2010-BarkeshliWen}.

\subsection{Dirac formulation of electrons in the lowest Landau level}
To start, we consider the problem of interacting nonrelativistic electrons and show, following Son~\cite{Son15}, that this problem is equivalent to Dirac electrons once the LLL limit is taken.
The Lagrangian for interacting electrons is given by
\be
\mathcal L = i \psi^\dag D_0 \psi - \frac{1}{2m_{e}} |D_i\psi|^2 + \frac{gB}{4m_{e}} \psi^\dag \psi - V_{\rm int}(|\psi|^2)\,,
\ee
where $g$ is the $g$ factor, $V_{{\rm int}}(|\psi|^2)$ is the interaction potential, and $D_\mu = \p_\mu - iA_\mu$ with $A_{\mu}$ the external electromagnetic vector potential corresponding to the magnetic field $B$. When $g{=}2$, we can take the limit $m_{e} \rightarrow 0$ to confine the electrons to the LLL. To do so, we re-write the Lagrangian as
\be
\mathcal L = i \psi^\dag D_0 \psi - \frac{1}{2m_{e}} (D_x - i D_y)\psi^\dag(D_x+iD_y)\psi - V_{\rm int}(|\psi|^2)\,.
\ee
Next, we introduce the Hubbard-Stratonovich field $\chi$ as follows:
\begin{multline}
\mathcal L = i \psi^\dag D_0 \psi + i \psi^\dag(D_x - i D_y)\chi + i \chi^\dag(D_x + iD_y)\psi 
\\ + 2m_{e} \chi^\dag \chi - V_{{\rm int}}\,.
\end{multline}
The limit $m_{e} \rightarrow 0$ is now smooth, and we can discard the mass term for $\chi$. 

Finally, we introduce the LLL Dirac electron 
\be
\Psi_e = 
\begin{pmatrix}
\psi 
\\
\chi
\end{pmatrix},
\ee
in terms of which the Lagrangian is given by
\be \la{eq_DiracE}
\mathcal L = i \bar{\Psi}_{e} \gamma^\mu D_\mu \Psi_e - V_{{\rm int}}\,. 
\ee
When dealing with relativistic degrees of freedom, we have to be careful with the filling fraction. Indeed, if $\psi$ forms a FQH state at filling $\nu_{\rm NR}$, then $\Psi_{e}$ must form a FQH state at filling $\nu{=}\nu_{\rm NR} - 1/2$~\cite{Son15, nguyen2017exact}.

\subsection{Parton construction}

Now we are ready to perform the parton construction. Note that the theory [see Eq.~\eqref{eq_DiracE}] is already projected to the LLL. Our main concern is to preserve this property. We assume that the nonrelativistic electron $\psi$ forms a Jain state at $\nu_{\rm NR} {=}n/(2pn{+}1)$. Then, the relativistic LLL electron $\Psi_e$ is at $\nu {=} n/(2pn{+}1){-}1/2$. We represent the LLL Dirac electron as a product of \emph{two} partons, one fermionic, $\Psi_p$, and one bosonic, $\phi$, as
\be
\Psi_e = \Psi_p \cdot \phi\,.
\ee
The Dirac field $\Psi_p$ has electric charge $q_1 {=} 1/p$ (we set the electron charge to unity), and $\phi$ has the electric charge $q_2 {=} 1 {-} 1/p$, so that $q_1 {+} q_2 {=} 1$ and the charge of the electron is recovered. We have to gauge the relative U(1) between $\Psi_p$ and $\phi$, which leaves the electron operator invariant to ensure that no unphysical states are introduced. This is accomplished by using a U(1) gauge field $\alpha$ to be introduced shortly.

After we introduce the partons, we must choose a mean-field state for each parton. We assume that $\Psi_p$ forms a Jain state at filling $\nu {=}n/(2n{+}1){-} 1/2$ and $\phi$ forms a Laughlin state at filling $\nu {=}1/[2(p{-}1)]$. This results in the Lagrangian
\begin{multline}
\mathcal L = i \bar{\Psi}_{p}\gamma^\mu\left[ \p_\mu - (i q_1 A_\mu-i\alpha_\mu) \right] \Psi_p - V_{{\rm int}}(\Psi_p)
\\
+ i \phi^\dag \left( \p_0 - i q_2 A_0-i\alpha_0 + \left|\p_i - iq_2 A_i - i \alpha_i \right|^2 \right) \phi - V_{{\rm int}}(\phi)\,,
\end{multline}
where $V_{\rm int}$ describes the residual interaction between the partons allowing them to form the aforementioned mean-field states.

We analyze this mean-field using an effective theory that respects the LLL projection. To do that, we dualize the fermionic parton and map the parton Jain state of $\ \Psi_p$ to an integer quantum Hall (IQH) state of the \emph{dual Dirac parton} $\Psi_d$ at filling $\nu {=} n{+}1/2$. 

The effective Lagrangian for $\nu {=}n/(2pn+1)$ is
\begin{multline} \la{eq_parton_theory}
\mathcal L = i \bar{\Psi}_d \gamma^\mu \left( \p_\mu - i a_\mu\right) \Psi_d - \frac{1}{4\pi} (q_1 \tilde A-\alpha) \wedge d a 
\\
+ \frac{1}{8\pi} (q_1 \tilde A-\alpha) \wedge d(q_1 \tilde A-\alpha) 
\\
+ i \phi^\dag \left( \p_0 - i q_2 A_0-i\alpha_0 + \left|\p_i - iq_2 A_i - i \alpha_i \right|^2 \right) \phi - V_{{\rm int}}(\phi)\,,
\end{multline}
where $a$ is a U(1) gauge field \cite{Son15} and we also enforce the massless Galilean invariance, which is generally present in the LLL~\cite{Golkar16}. To leading order, this requires an inclusion of the spin connection $\omega$ with the vector potential as follows:
\be
\tilde A = A + \frac{p}{2} \omega\,.
\ee
The Dirac field $\Psi_d$ represents the electrically neutral composite Dirac parton and $\phi$ is the bosonic parton forming the $\nu {=} 1/[2(p{-}1)]$ Laughlin state. The two interact via the gluing U(1) gauge field $\alpha$. Equation~\eqref{eq_parton_theory} is the central result of this section. The collective modes are spin-$2$ excitations of $\Psi_d$ (or, alternatively, $\Psi_p$) and spin-$2$ excitations of $\phi$. 
In the remainder of the section, we argue that Eq.~\eqref{eq_parton_theory} is the correct effective theory of LLL electrons forming a Jain state at $\nu{=}n/(2pn{+}1)$. 

\subsection{Response theory}
We would like to make sure that the effective theory given in Eq.~\eqref{eq_parton_theory} gives the correct Hall conductivity and Wen-Zee shift. As a side result, we derive theories introduced in Refs.~\cite{Goldman18, wang2019dirac}. 

First, we integrate out the high-energy bosonic parton $\phi$, which generates the following contribution to the effective Lagrangian:
\begin{multline}
\mathcal L_\phi = \frac{1}{4\pi} \frac{1}{2(p-1)} (q_2 A + \alpha) d (q_2 A+\alpha) 
\\
+ \frac{2(p-1)}{4\pi} \frac{1}{2(p-1)}(q_2A + \alpha)d\omega \,.
\end{multline}

Second, we integrate out the gluing gauge field $\alpha$. The equation of motion for $\alpha$ takes the form
\be
\alpha = -\frac{p-1}{p}\left( a + \frac{1}{2}\omega \right) \,,
\ee
which leads to the following effective Lagrangian
\begin{multline}\la{eq_L_noalpha}
\mathcal L = i \bar{\Psi}_d \gamma^\mu \left( \p_\mu - i a_\mu\right) \Psi_d - \frac{p-1}{8\pi p}ada - \frac{1}{4\pi p} Ada + \frac{1}{8\pi p} AdA
\\
- \frac{2p-1}{8\pi p} ad\omega + \frac{1}{4\pi} \frac{2p-1}{2p}Ad\omega\,. 
\end{multline}
The first line in Eq.~\eqref{eq_L_noalpha} is the Lagrangian studied in Refs.~\cite{Goldman18, wang2019dirac}. The second line is the coupling to the spin connection.

To obtain the linear response functions, we need to integrate out the composite Dirac parton $\Psi_d$. To do so, we follow the method outlined in Refs.~\cite{Golkar14, Golkar16}. The effective Lagrangian generated from integrating out $\Psi_d$ is 
\be \la{eq_Leff_d}
\mathcal L_{\Psi_d} = \frac{1}{4\pi}\left( n+ \frac{1}{2}\right) ada + \frac{n(n+1)}{4\pi} ad\omega \,.
\ee
Integrating out $a$ from Eqs.~\eqref{eq_Leff_d} and \eqref{eq_L_noalpha} gives the induced Lagrangian that encodes the Hall conductivity and the shift
\be
\mathcal L = \frac{n}{2pn+1} \frac{1}{4\pi} AdA + \frac{n (n+2p)}{2pn+1} \frac{1}{4\pi} Ad\omega,
\ee
which confirms the correct Hall conductivity and the shift for the Jain state at $\nu{=}n/(2pn{+}1)$.

\subsection{Negative Jain series and reflection symmetry}

The states in the negative Jain series at $\nu{=}n/(2pn{-}1)$ are obtained by applying the particle-hole transformation to the Dirac parton
\be
n+\frac{1}{2} \rightarrow - \left(n + \frac{1}{2}\right)\,.
\ee
The resulting response action is then simply
\be
\mathcal L = \frac{1}{4\pi} \frac{n+1}{2p(n+1) - 1} AdA + \frac{n [-(n+1) + 2p]}{2pn+1} \frac{1}{4\pi} Ad\omega \,,
\ee
which is just the negative Jain series with $n'{=}n{+}1$.

This particle-hole transformation is \emph{not} the particle-hole symmetry of the LLL electrons near half filling. This transformation does not have to preserve the spectrum of the original electron problem. It does, however, preserve the spectrum of the Dirac partons, since they are close to half filling and it does not affect the spectrum of $\phi$ at all. This may explain why enforcing the reflection symmetry discussed in Ref.~\cite{Goldman18} is the right strategy. This transformation changes the chirality of the parton state, suggesting that the lower-energy collective modes for positive and negative Jain series have different chiralities, while the high-energy collective mode has the same chirality.

\subsection{Collective modes}
Next, we discuss the collective modes. Qualitatively, the presence of two collective modes is explained as follows. At small wave numbers, we can excite either the quadrupole mode of the dual Dirac parton $\Psi_d$ or the quadrupole mode of the bosonic Laughlin parton field $\phi$. To show this rigorously, we start with Eq.~\eqref{eq_parton_theory}. 

First, we analyze the bosonic Laughlin state. This analysis follows the original bimetric discussion in Ref.~\cite{Gromov17}. The putative effective bimetric theory that includes the GMP mode takes the form
\begin{multline}
\mathcal L = -\frac{2(p-1)}{4\pi} \beta d \beta - \frac{1}{2\pi} \left(\frac{p-1}{p} A + \alpha\right)d\beta \\
- \frac{s}{2\pi}\omega d \beta - \frac{\varsigma}{2\pi} \hat{\omega} d \beta - H[\hat{g}]\,,
\end{multline}
where $\beta$ is the gauge field in the U(1) Chern-Simons theory describing the topological order and $s$ and $\varsigma$ are phenomenological coefficients to be fixed shortly, while $\hat{g}$ and $\hat \omega$ are the dynamical degrees of freedom describing the fluctuations of the spin-$2$ degrees of freedom \cite{Gromov17}. The Hamiltonian $H[\hat g]$ can be chosen to be $H {=} m_{g}\Tr(g \hat{g}^{{-}1})$ \cite{lapa2019geometric}, where $m_{g}$ is the mass of $\hat{g}$ . 
Next, we integrate out the topological field $\beta$. The equation of motion
\be
\beta = -\frac{1}{2} \left[\frac{1}{p} A + \frac{1}{p-1}\alpha + \frac{s}{p-1}\omega + \frac{\varsigma}{p-1}\hat{\omega} \right]
\ee
leads to the Lagrangian
\begin{multline} \la{eq_BM}
\mathcal L = \frac{1}{4\pi} \frac{1}{2(p-1)}\left[\frac{p-1}{p} A + \alpha\right] d \left[\frac{p-1}{p} A + \alpha\right] 
\\
+ \frac{1}{4\pi}\frac{2s}{2(p-1)} \left(\frac{p-1}{p}A + \alpha\right) d \omega 
\\
+ \frac{1}{4\pi}\frac{2\varsigma}{2(p-1)} \left(\frac{p-1}{p}A + \alpha\right) d \hat{\omega}- H[\hat g].
\end{multline}
When $m_{g}$ is large, $\hat{g} {=} g$ and $\hat{\omega} {=} \omega$. In this limit, we get the Lagrangian of Eq.~\eqref{eq_L_noalpha} studied in the previous section. That Lagrangian leads to the correct shift $\mathcal S$. Thus we get a constraint (same constraint as in bimetric theory)
\be
\mathcal S = 2(s + \varsigma)\,.
\ee
To fix $\varsigma$, we need to compute the static structure factor. To do this, we add the first line in Eq.~\eqref{eq_L_noalpha} to Eq.~\eqref{eq_BM} and integrate out $\alpha$ to obtain
\begin{multline}\la{eq_final_L}
\mathcal L = i \bar{\Psi}_c \gamma^\mu\left(\p_\mu - i a_\mu\right)\Psi_d - \frac{1}{8\pi}\left(1 - \frac{1}{p}\right)ada - \frac{1}{4\pi p}Ada 
\\
+ \frac{1}{4\pi}\frac{2\varsigma}{2p}(A-a)d\hat \omega 
+ \frac{1}{8\pi p}AdA + \frac{1}{4\pi}\frac{1+2s}{2p}(A-a)d\omega\,.
\end{multline}
This is the final Lagrangian describing the parton Dirac Fermi liquid coupled to the bimetric theory of the Laughlin state. The last term can be dropped by turning off the background geometry. The coefficient $\varsigma$ still has to be fixed. When $\hat \omega {=} \omega$ and $2(s{+}\varsigma){=}2(p{-}1)$ this Lagrangian gives back Eq.~\eqref{eq_L_noalpha}.

Equation~\eqref{eq_final_L} is not exactly the same theory as in Ref.~\cite{Nguyen2021}, because there is a coupling between Dirac and bimetric theory due to the term $[1/(4\pi)](2\varsigma/(2p)(A{-}a)d\hat \omega)$. In Appendix~\ref{app: EFT}, we perform a detailed calculation of the projected static structure factor, which reveals that $\varsigma {=} p{-}1$ and $s{=}0$, in agreement with Ref.~\cite{Nguyen2021}. Upon reflection, this is not surprising. To the leading order in gradients, no term can couple the spin-$2$ fluctuation of the Fermi surface to the parton geometric degree of freedom. The leading term to induce such coupling takes the form of a gravitational Chern-Simons and should affect $q^6$ correction to the projected static structure factor.

\section{Wave function for the parton collective mode}
\label{sec: parton_wf}

In this section, based on the microscopic parton construction, we propose a trial wave function to capture the parton mode. As we show below, a description of this collective mode lies beyond the purview of the CF theory. We start with a brief introduction to the microscopic version of the parton theory of the FQH effect (FQHE).

Jain generalized his CF theory to introduce the parton theory of FQHE~\cite{Jain89b}, where the partons are nonrelativistic fermions that occupy IQH states (with this definition, CF states are a subset of parton states, but, throughout this work, we reserve the phrase parton state to denote a non-CF state). The parton theory constructs trial wave functions for an FQHE state as a product of IQH wave functions. Until very recently, none of the parton states were found to be relevant to describe experimentally observed states. In the past few years, there has been a resurgence~\cite{Wu17, Balram18, Balram18a, Balram19, Kim19, Faugno19, Balram20, Balram20a, Faugno20a, Faugno20b, Balram21, Balram21a, Balram21b, Dora22} in the parton theory whereby it now appears that viable candidate parton states can be constructed for all FQH plateaus observed to date that do not fit the paradigm of weakly interacting composite fermions (see Refs.~\cite{Balram20a, HalperinJainBook} for a summary of these works). These works should be viewed as complementary to ours, since they almost exclusively focus on ground states whereas we are interested in excitations that lie high up in the spectrum. Furthermore, they deal with states that lie outside the Jain sequence while in this article we look at only states that reside in the Jain series.

We note here that in the microscopic version of the parton theory, IQH states of nonrelativistic fermions are the building blocks, and, in this section, we follow this convention. This should be contrasted with the description given in the previous section where the partons were either Dirac fermions or bosons and form FQH states. Nevertheless, as we show below, the wave functions we propose can be cast in a form that closely resembles the corresponding states that we construct in the previous section using a field-theoretic treatment. 

We first begin by introducing Jain's CF ansatz to represent the FQHE ground states of our interest~\cite{Jain89}. The CF ground state wave function at $\nu{=}n/(2pn{\pm}1)$ is given by
\begin{equation}
\Psi^{\rm Jain}_{n/(2pn\pm 1)} = \mathcal{P}_{\rm LLL} \Phi^{2p}_{1}\Phi_{\pm n},
\label{eq: Jain_wf}
\end{equation}
where $\Phi_{n}$ is the Slater determinant state of $n$-filled LLs of electrons (with $\Phi_{{-}n} {=} [\Phi_{n}]^{*}$) and $\mathcal{P}_{\rm LLL}$ implements projection to the LLL as is appropriate in the high magnetic field limit. The Laughlin-Jastrow factor $\Psi^{\rm Laughlin}_{1/(2p)}{=}\Phi_{1}^{2p}$~\cite{Laughlin83} attaches $2p$ vortices to each of the electrons to turn them into composite fermions. Thus, the FQHE ground state of electrons at $\nu{=}n/(2pn{\pm}1)$ can be viewed as the $\nu^{*}{=}n$ IQH state of composite fermions carrying $2p$ vortices. Similarly, the excitations of the FQH states can be constructed by appealing to this mapping to IQH states. In particular, the lowest-lying neutral excitation termed the CF exciton (CFE) is obtained by replacing $\Phi_{n}$ in Eq.~\eqref{eq: Jain_wf} by $\Phi^{\rm ex}_{n}$, where an exciton (ex) is a particle-hole pair with the hole residing in the LL indexed by $n{-}1$ and the particle in the LL indexed by $n$. The Jain wave functions provide an excellent description of the Coulomb states obtained from exact diagonalization in the LLL~\cite{Jain07, Balram13, Zuo20, Balram20a, Balram20}. A nice feature of these wave functions is that they can be evaluated in~\emph{real space} for hundreds of electrons using the Jain-Kamilla (JK) method of projection~\cite{Jain97, Jain97b, Jain07, Moller05, Davenport12, Balram15a}. 

The Jain wave function can be interpreted as being composed of $(2p{+}1)$ partons~\cite{Jain89b} with $2p$ partons, each of charge $e_{1}{=}(-e)n/(2pn{\pm}1)$, forming a $\nu{=}1$ IQH state and one parton, of charge $e_{\pm n}{=}{\pm}({-}e)/(2pn{\pm}1)$, forming a $\nu{=}{\pm}n$ IQH state. The low-energy excitations of the $\nu{=}n/(2pn{\pm}1)$ are constructed from excitations in the $\nu{=}\pm n$ IQH state, since $|e_{{\pm}n}|\leq |e_{1}|$ (equality holds only for $n{=}1$) which results in a smaller Coulomb penalty. This is the reason why the low-lying CFE mode is made up of a particle-hole pair in the $\Phi_{\pm n}$ factor. However, high-energy modes can arise from excitations in the $\nu{=}1$ IQH state. As we show below, it is precisely an excitation of this kind, namely, a particle-hole pair in the Jastrow factor, that forms the parton mode. This argument shows why the parton mode is absent in the Laughlin fractions. Although the $\nu{=}1/(2p{+}1)$ Laughlin state is also made up of $(2p{+}1)$ partons, the partons are all of the same kind; i.e., they all form a $\nu{=}1$ IQH state. Thus, the particle-hole pair excitation can be created only in a $\Phi_{1}$ factor, and, thus, we do not end up with any additional modes besides the CFE one. More generally, any Jain state can host at most two spin-$2$ modes since, it harbors only two kinds of partons. 

With this background in place, we next introduce the wave functions for the CF exciton and parton modes. The wave function for the CFE mode is given by $\Psi^{\rm CFE}_{n/(2pn\pm 1)} {=} \mathcal{P}_{\rm LLL} \Phi^{2p}_{1} \Phi^{\rm ex}_{\pm n}$. In contrast, the wave function for the parton mode, for $n,p{\geq}2$, is given by 
\begin{eqnarray}
\Psi^{\rm parton~mode}_{n/(2pn\pm 1)} &=& \mathcal{P}_{\rm LLL} \Phi^{\rm ex}_{1}\Phi^{2p-1}_{1} \Phi_{\pm n} \nonumber \\
&\sim& \left( \mathcal{P}_{\rm LLL} \Phi^{\rm ex}_{1}\Phi^{2p-3}_{1} \right) \times \left( \mathcal{P}_{\rm LLL} \Phi^{2}_{1}\Phi_{\pm n} \right) \nonumber \\
&=&\Psi^{\rm CFE}_{1/[2(p-1)]} \Psi^{\rm Jain}_{n/(2n\pm 1)},
\label{eq: parton_mode}
\end{eqnarray}
where $ \Psi^{\rm CFE}_{1/[2(p{-}1)]} {=} \Psi^{\rm CFE}_{1/(2p{-}1)}/\Phi_{1}$. The wave function for the parton mode given in Eq.~\eqref{eq: parton_mode} is the central result of this section. In Eq.~\eqref{eq: parton_mode}, the $\sim$ sign indicates that the projection to the LLL is carried out separately on two parts of the wave function in order to facilitate its evaluation for large systems. We expect that such details of the projection do not affect the qualitative nature of the state and lead to only minor differences in its quantitative properties~\cite{Balram15a, Balram16b}. Note that the CFE mode can also be projected in a similar fashion which results in the wave function $\Psi^{\rm CFE}_{n/(2pn\pm 1)} \sim\Psi_{1/[2(p-1)]} \times \Psi^{\rm CFE}_{n/(2n\pm 1)}$. Written in this form as a product of a bosonic Laughlin state times a Jain state (with or without excitations), these wave functions closely resemble the analogous states constructed in the previous section using field-theoretic techniques. The wave function of Eq.~\eqref{eq: parton_mode} also demonstrates that for $p{>}2$, the additional Jastrow factors do not lead to any extra parton modes since $\mathcal{P}_{\rm LLL}\Phi^{2p}_{1}\Phi^{\rm ex}_{1}{=}\Phi^{2}_{1}\mathcal{P}_{\rm LLL}\Phi^{2p-2}_{1}\Phi^{\rm ex}_{1},~\forall p{\geq2}$~\cite{Balram15a}.

\begin{figure}[bhtp]
		\includegraphics[width=1\linewidth]{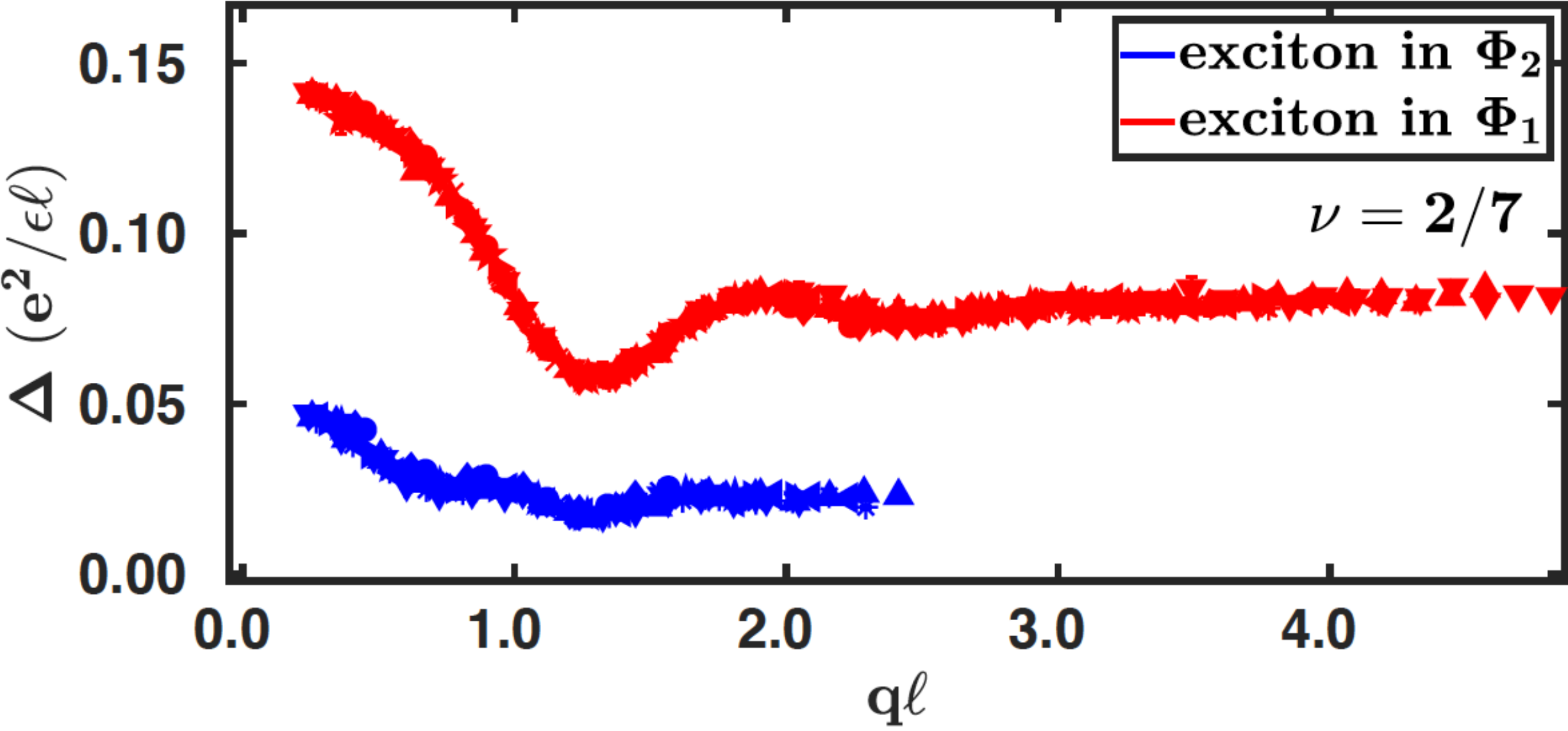}
		\caption{Coulomb energies of the CF exciton and parton modes at $\nu{=}2/7$. Different system sizes are plotted with different symbols with the smallest system with $N{=}12$ electrons and the largest with $N{=}40$. In the long-wavelength limit, the high-energy (red) mode extrapolates to an energy of approximately 0.15, while the low-energy (blue) mode extrapolates to approximately 0.05 in Coulomb units. }
		\label{fig: CFE_parton_modes_2_7}
\end{figure}

In Fig.~\ref{fig: CFE_parton_modes_2_7}, we show the energies of the CF exciton and parton modes at $\nu{=}2/7$ (see Appendix~\ref{app:2_9_Jain} for the dispersions of the two modes at $\nu{=}2/9$). These energies are obtained as Coulomb expectation values of the wave functions given above in the spherical geometry~\cite{Haldane83}. Throughout this work, we quote energies in units of $e^2/(\epsilon\ell)$, where $\epsilon$ is the dielectric constant of the background host and $\ell{=}\sqrt{\hbar/(eB)}$ is the magnetic length at magnetic field $B$. In Fig.~\ref{fig:sq} below, we  show that the energies of the two modes in the long-wavelength limit coincide with the part of the spectrum in which the dynamical structure factor has maximal support. 

Next, we turn to the chirality of the modes. The low-energy CFE mode at $2/9$ and $2/7$ stem from the $2/5$ and $2/3$ parts of the wave function, respectively. Therefore, the CFE mode at $2/9$ and $2/7$ has the same chirality as that of the GMP mode in the $2/5$ and $2/3$ states, respectively. On the other hand, the high-energy mode arises from the CFE of the 1/2 Laughlin state, and, therefore has the same chirality as that of the GMP mode of the 1/2 Laughlin state (in our convention, the GMP mode of the Laughlin state has a negative chirality). These observations allow us to predict that (i) the chirality of the two modes is the \emph{same} at $2/9$ and is negative and (ii) the chirality of the two modes at $2/7$ is \emph{opposite} to each other with the CFE mode having a chirality ${+}$ and the parton mode having a chirality of ${-}$. At $\nu{=}n/(2pn{+}1)$, the effective magnetic field seen by all the partons is in the same direction as the electrons. In contrast, at $\nu{=}n/(2pn{-}1)$, the parton forming the $n$-filled LL IQH state experiences a magnetic field that is opposite to that of the electrons (the other partons see a magnetic field that is parallel to that of the electrons). Thus, the chirality of a mode is simply determined by the sign of the effective magnetic field seen by the parton hosting the excitation. We confirm these predictions by explicitly evaluating the chiral dynamical response functions in Fig.~\ref{fig:vmn} below.

Finally, we note that a different version of the wave function for the modes can be obtained by following the GMP construction, i.e., by acting with the LLL projected density operator on the ground state. However, the density operator selectively acts only on a particular part of the wave function: In essence, we apply the density operator $\rho(\vec{q})$ onto the unprojected Jain wave function but project it into the LLL in different ways. These lead to the following wave functions for the two modes: (i) the primary low-energy mode $\Psi^{\rm GMP}_{n/(2pn\pm 1)}\sim \Psi_{1/[2(p-1)]} \times \Psi^{\rm GMP}_{n/(2n\pm 1)}$, and (ii) for $n, p{\geq}2$, the secondary high-energy mode $\Psi^{\rm GMP}_{1/[2(p-1)]} \times \Psi_{n/(2n\pm 1)}$, where $\Psi^{\rm GMP} {=} \bar{\rho}(\vec{q})\Psi$ with $\Psi$ being the ground state wave function and $\bar{\rho}(\vec{q})$ the LLL-projected density operator. Unlike the parton and CF-exciton wave functions given above, these GMP-based wave functions are expected to work well only in the long-wavelength limit~\cite{Yang12b, Repellin2014}. 

\section{Dynamical response functions} 
\label{sec: dynamical_sf}
In this section, we compute the dynamical structure factor and chirality-resolved spectral functions, which allows one to directly probe the existence of parton modes in unbiased microscopic simulations using exact diagonalization. In order to identify collective modes, we compute dynamical response functions $I(E)$ defined by
\begin{eqnarray}\label{eq:sq}
I (E) = \sum_n |\langle E_n | \hat O |0\rangle |^2 \delta(E_n - E_0 - E),
\end{eqnarray}
where $\hat O$ is an operator (to be specified below) and the sum runs over all eigenstates $|E_n\rangle$ with energies $E_n$, with $|0\rangle$ denoting the ground state at energy $E_0$. We model the FQH system by following the standard procedure~\cite{Haldane83, ChakrabortyBook} by placing $N$ electrons on a compact domain, such as the sphere or torus, threaded by $N_\phi$ magnetic flux quanta. In the main text, we primarily focus on the FQH state at $\nu{=}2/7$, which occurs at a shift $\mathcal{S}{=}2$ and belongs to the secondary Jain sequence of CF states (see appendixes for results at other filling factors). 

Before presenting results for spectral functions, we note that the evaluation of Eq.~\eqref{eq:sq} potentially requires summing over a large number of eigenstates. For this, we make use of the rotational symmetry on the sphere and target about $40{-}60$ lowest eigenstates in each angular momentum $L$ sector using the Lanczos method. We further ``polish" the eigenstates by running Lanczos iterations using the $\mathbf{L}^2$ operator, to ensure that the eigenstates are rotationally invariant to machine precision.

\begin{figure}[htb]
  \centering
  \includegraphics[width=0.9\linewidth]{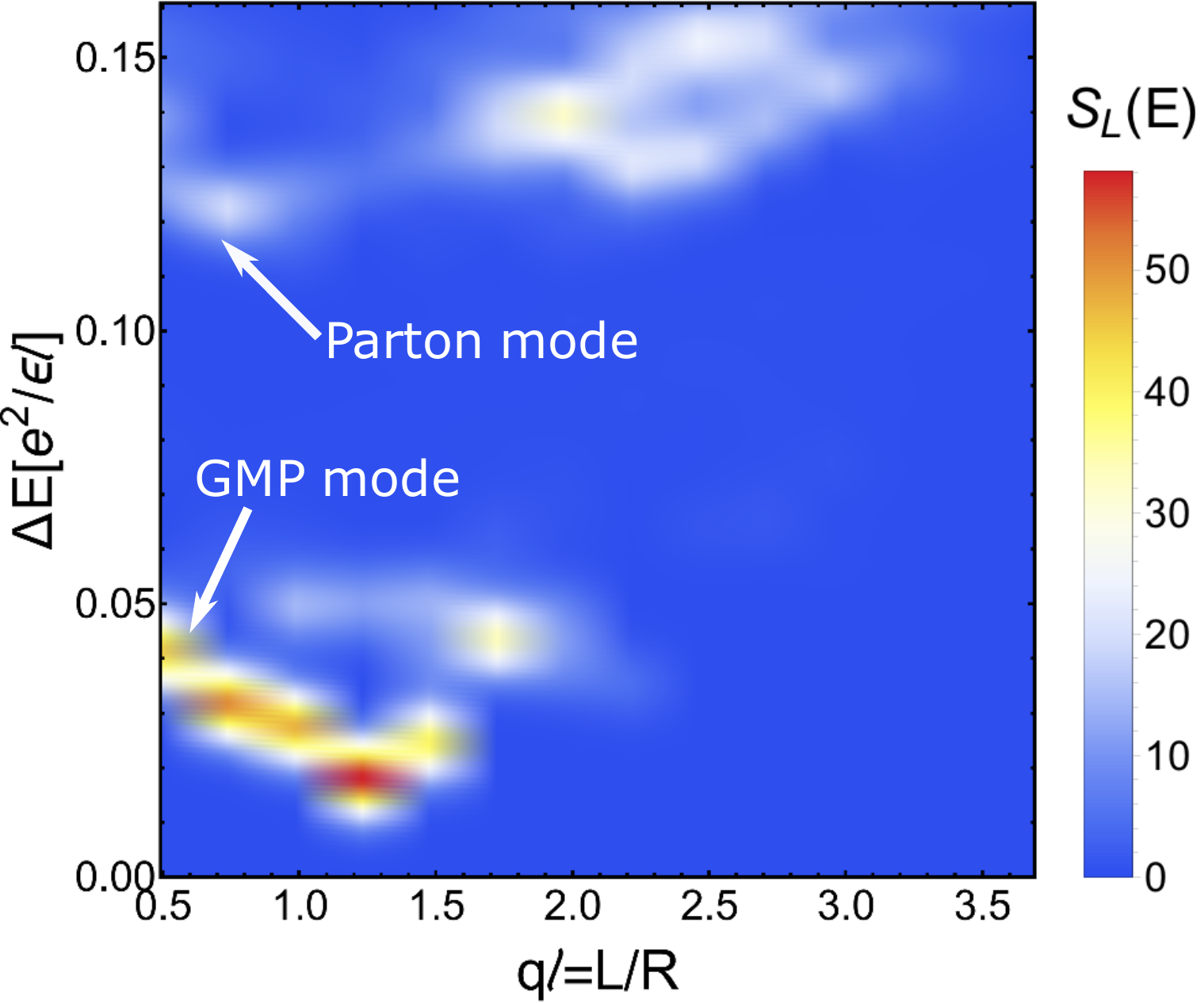}
  \caption{Dynamical structure factor $S_L(E)$ at $\nu{=}2/7$ for $N{=}10$ electrons on the sphere, plotted as a function of linear momentum $q\ell{=}L/R$, where $L$ is the angular momentum and $R$ is the sphere radius, and energy $E$ is measured relative to the ground state energy. The two modes are indicated by arrows. }
  \label{fig:sq}
\end{figure}
Choosing $\hat O$ to be the projected density operator, $\bar{\rho}_{L}$, the spectral function in Eq.~\eqref{eq:sq} is nothing but the dynamical structure factor~\cite{HeSimonHalperin1994, HePlatzman1994, Platzman96}. This dynamical structure factor $S_L(E)$ is evaluated for the $\nu{=}2/7$ state and shown in Fig.~\ref{fig:sq}. To facilitate comparison with the infinite plane geometry, $S_L(E)$ is plotted as a function of the planar momentum $q\ell {=} L/R$, where $R{=}\ell\sqrt{N_{\phi}/2}$ is the radius of the sphere. We approximate the delta function in Eq.~\eqref{eq:sq} by a Gaussian with width $5{\times}10^{-3}$. We observe that most of the contribution to $S_L(E)$ comes from the mode at the energy slightly below $E{\approx}0.05$, which is identified with the GMP mode below. However, there are weaker yet clearly visible signatures of an additional collective mode around the energy $E{\approx}0.13$. Note that we can also observe some contribution to $S_L(E)$ from states at large momenta. In Ref.~\cite{Platzman96}, those are interpreted as local distortions of the quasiparticle-quasihole droplets. 

To identify the collective modes more accurately, we perform a finer characterization of the dynamical response by choosing $\hat O$ in Eq.~\eqref{eq:sq} to be one of the anisotropic pseudopotentials $V_{m,s}^\sigma$ with a given chirality $\sigma{=}\pm$~\cite{KunYang2016, YangGeneralizedPP, Liu18, Liou19}. This allows us to probe systematically how the system responds to a particular type of metric deformation. For example, mass anisotropy leads to a predominantly quadrupolar metric deformation~\cite{YangGeneralizedPP,YangTilted}; hence, we can fix $s{=}2$. On the other hand, $m$ contains information about the clustering properties of the electrons. In the absence of anisotropy, $m$ becomes the relative angular momentum which is associated with the standard Haldane pseudopotential~\cite{Haldane83}. Finally, for the definition of chirality, we adopt the convention stated in the previous section; i.e., $\sigma{=}{-}$ is the chirality of the GMP mode of the Laughlin state at $\nu{=}1/3$~\cite{Liou19}.

\begin{figure}[thb]
  \centering
  \includegraphics[width=\linewidth]{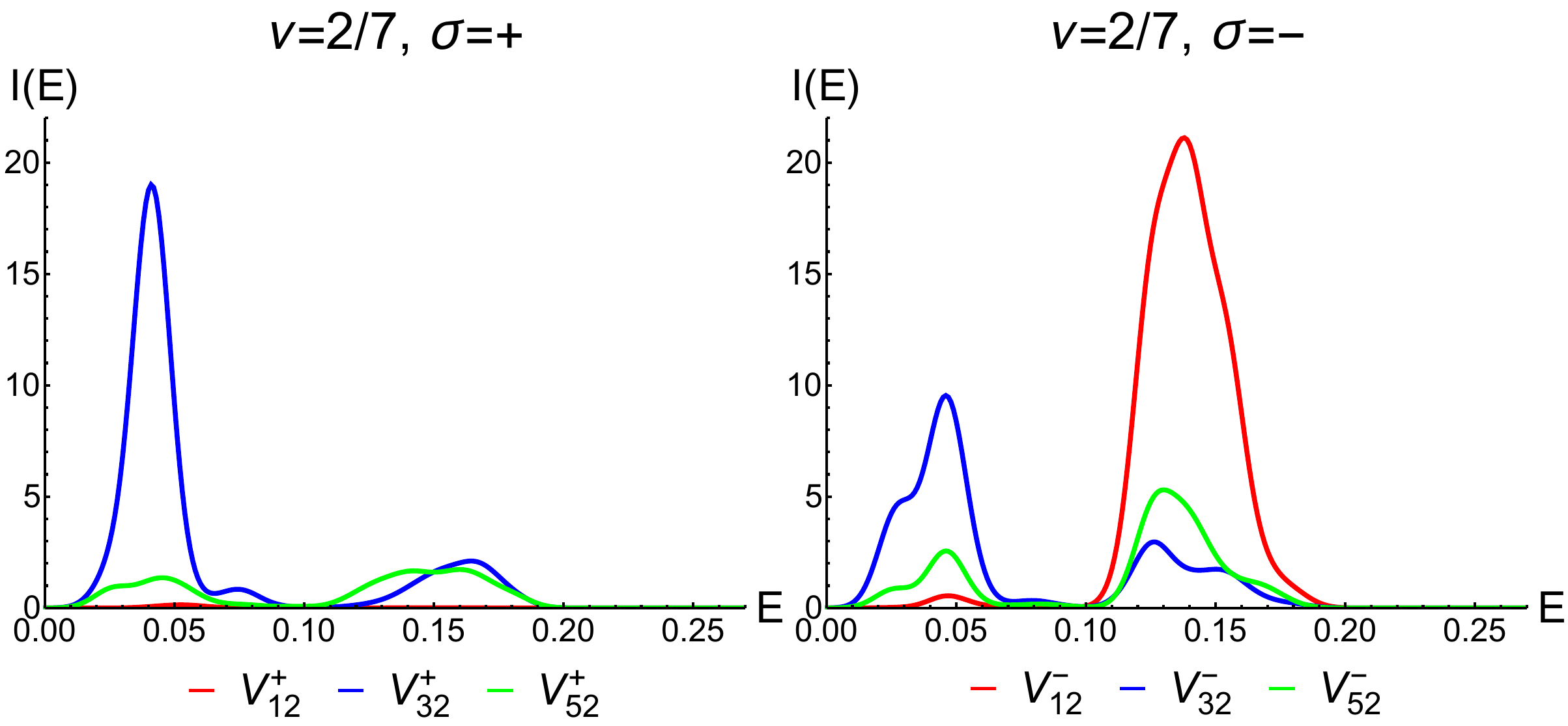}
  \caption{Spectral functions for the generalized pseudopotentials $V_{m,s}^\sigma$ with chirality $\sigma$ at $\nu{=}2/7$. Two dominant peaks, at energies $E\approx 0.14$ and $E{\approx}0.04$, have \emph{opposite} chiralities. Data are for $N{=10}$ electrons on the sphere. }
  \label{fig:vmn}
\end{figure}
Anisotropic spectral functions for the $\nu{=}2/7$ state are shown in Fig.~\ref{fig:vmn} for $m{=}1,3,5$ and $\sigma{=}\pm$ (with $s{=}2$). Since anisotropic pseudopotentials break the $\mathbf{L}^2$ symmetry, these spectral functions are evaluated in the $L_z{=}0$ sector and they are only a function of the energy $E$. Moreover, the spectral functions are normalized such that the integrated $I(E)$, summed over both chirality sectors, is approximately equal to 1. From Fig.~\ref{fig:vmn}, we conclude that $\nu{=}2/7$ has two dominant peaks in its spectral response, supporting the existence of two collective modes in the long-wavelength limit. The chirality of these modes is \emph{opposite} at $\nu{=}2/7$ but can be the same at other fillings, e.g., $\nu{=}2/9$, as we shown in Appendix~\ref{app:2_9_Jain}. 

The spectral function peaks in Fig.~\ref{fig:vmn} are naturally accounted for by considering the parton wave function of Eq.~\eqref{eq: parton_mode} and that of the usual CF exciton. At $\nu{=}2/7$, the high-energy mode is $\Psi^{\rm Jain}_{2/3}\times \Psi^{\rm CFE}_{1/2}$. When two particles approach each other, $\Psi^{\rm Jain}_{2/3}$ vanishes as the first power of their interparticle spacing $r$ while $\Psi^{\rm CFE}_{1/2}$ does not vanish, so overall the parton mode vanishes as $r$. Therefore, the high-energy mode is expected to have a peak in the $m{=}1$ spectral function. By contrast, the low-energy mode is $\Psi^{\rm CFE}_{2/3}\times \Psi^{\rm Laughlin}_{1/2}$. Now, $\Psi^{\rm CFE}_{2/3}$ vanishes as $r$ while $\Psi^{\rm Laughlin}_{1/2}$ vanishes as $r^{2}$, so overall the wave function vanishes as $r^{3}$ when two particles are brought close to each other. Thus, the low-energy mode has $m{=}3$. In general, at $\nu{=}n/(2pn\pm 1)$ with $n,p{\geq}2$, the high-energy mode has $m{=}1+2(p-2)$ [$1$ from the Jain state at $\nu{=}n/(2n\pm 1)$ and $2(p-2)$ from the CFE of the bosonic Laughlin state at $\nu{=}1/[2(p-1)]$], while the low-energy mode has $m{=}1+2(p-1)$ [$1$ from the CFE of the Jain state at $\nu{=}n/(2n\pm 1)$ and $2(p-1)$ from the bosonic Laughlin state at $\nu{=}1/[2(p-1)]$]. The spectral function shown in Fig.~\ref{fig:vmn} is consistent with these clustering properties. We note here that these clustering properties hold only approximately for eigenstates of the LLL-projected Coulomb interaction. Finally, we observe in Fig.~\ref{fig:vmn} that the energies corresponding to peaks in the spectral functions are in quantitative agreement with the variational estimates of the collective mode energies shown in Fig.~\ref{fig: CFE_parton_modes_2_7}.

\section{Quench dynamics} 
\label{sec: quench_dynamics}
In this section, we show that the multiple collective modes can be probed using a geometric quench~\cite{Liu18}. This quench targets the spin-$2$ degrees of freedom in the long-wavelength limit. One way to drive this type of quench is to suddenly introduce anisotropy in the electron's effective mass tensor, which can be experimentally implemented by tilting the magnetic field. It has been demonstrated that such mass anisotropy quenches indeed excite the spin-$2$ GMP modes of the Laughlin state~\cite{Liu18} and the bilayer Halperin state~\cite{Liu2021}. 

Our implementation of the geometric quench is conveniently performed in the torus geometry~\cite{ChakrabortyBook}. In momentum space, the FQH Hamiltonian is given by
$ H = \sum_{{\bf q}} \bar{V}_{\bf q} { \rho}^{\sigma}_{\bf q} { \rho}^{\sigma'}_{-{\bf q}}$, where $\rho_{\bf q}$ is the density operator projected to a Landau level and ${\bar V}_{\bf q}{=}(2\pi/|\mathbf{q}|) |F_{\bf q}|^2$ is the Fourier transform of the Coulomb interaction, dressed by the LL form factor $F_\mathbf{q}$. The form factor $F_{\bf q}{=}\exp[-g_m^{ab}q_a q_b \ell^2/4]$ is determined by the $2{\times}2$ mass tensor $g_m$~\cite{Haldane11} (with Einstein's summation convention implicit). In the isotropic case, we have $g_m{=}\id$, where $\id$ is the identity matrix. We initialize the quench by instantaneously changing $g_m$ from identity to ${\rm diag}\{\alpha,1/\alpha\}$ at time $t{=}0$, and we monitor the fidelity $F(t){=}|\langle\Psi(0)|\Psi(t)\rangle|$, i.e., the overlap between the initial state $|\Psi(0)\rangle$ and the evolved state $|\Psi(t)\rangle$. 

\begin{figure}[thb]
		\includegraphics[width=1\linewidth]{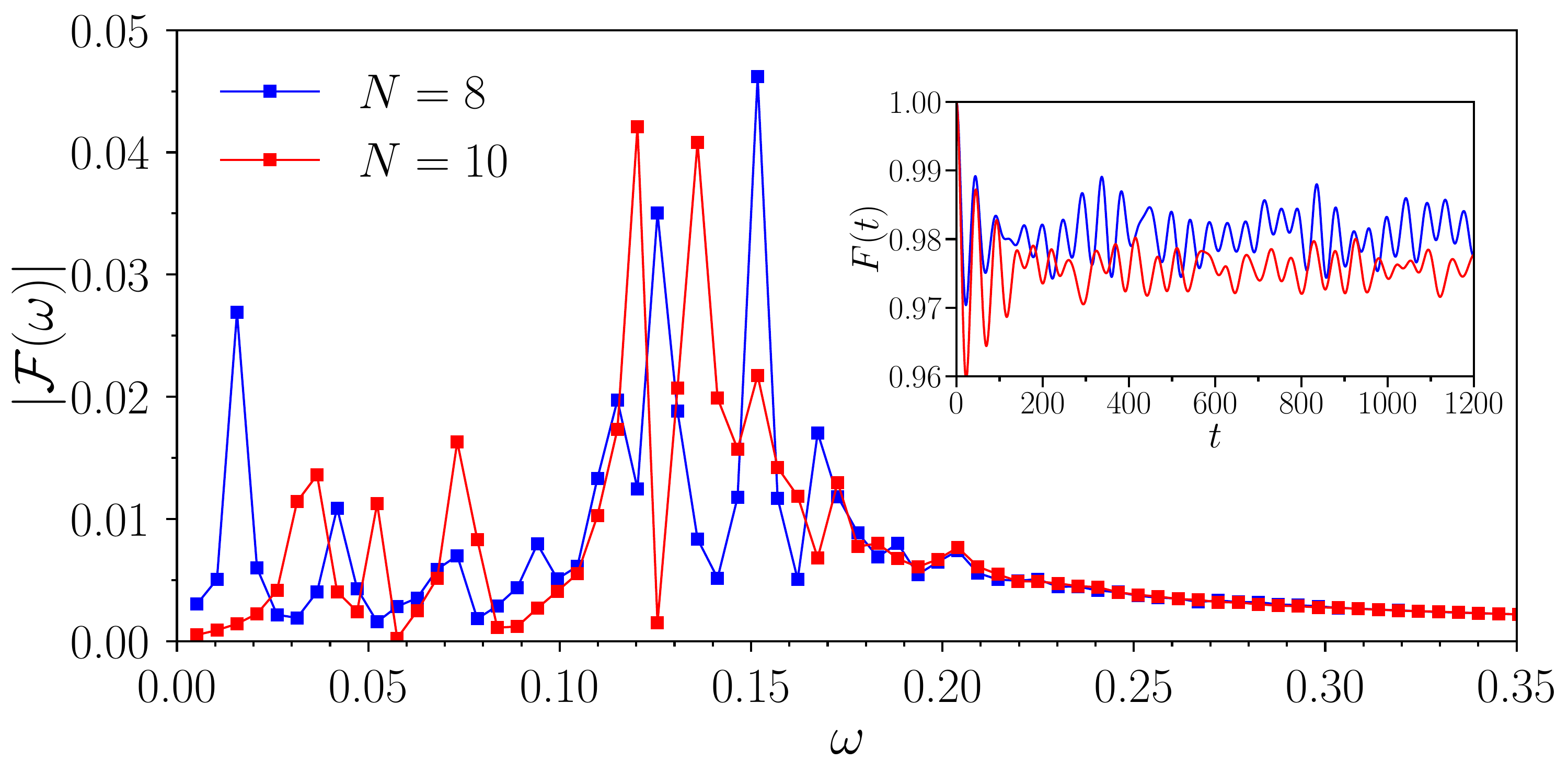}
		\caption{Fourier transform of the postquench fidelity (inset) for the mass anisotropy quench with $\alpha{=}1.3$. Data are for $N{=}8$ and $N{=}10$ electrons at $\nu{=}2/7$ on a torus with the square unit cell.}
		\label{fig:dynamics_2_7}
\end{figure}
The results for fidelity dynamics at $\nu{=}2/7$ are shown in Fig.~\ref{fig:dynamics_2_7}. The postquench fidelity $F(t)$ oscillates clearly with multiple frequencies (inset in Fig.~\ref{fig:dynamics_2_7}). To extract the dominant frequencies, we calculate the discrete Fourier transform $|\mathcal{F}(\omega)|$ of $F(t)$. We observe two groups of pronounced peaks in $|\mathcal{F}(\omega)|$, which are centered around frequencies $\omega{=}0.05$ and $\omega{=}0.13$, respectively. These dominant peaks within two narrow frequency windows are strong evidence of two spin-$2$ collective modes in the long-wavelength limit. Moreover, the energies of these two modes reflected in the quench dynamics are consistent with those estimated from the dynamical structure factor (Fig.~\ref{fig:sq}) and spectral functions (Fig.~\ref{fig:vmn}) on the sphere geometry.

\section{Discussion and conclusions}
\label{sec: discussion_conclusions}

In this paper, we argue that high-energy spectral properties of FQH states harbor evidence for the existence of partons -- the \emph{bona fide} quasiparticles of these strongly correlated topological quantum fluids. 
While at sufficiently low energies partons remain hidden to conventional probes, such as transport, their presence is revealed at high energies by exposing the FQH system to a geometric quench or acoustic wave absorption~\cite{KunYang2016}. The parton collective mode, identified in this work, emerges as the missing ingredient that ensures the consistency of the field-theoretic description of general Jain FQH states~\cite{Nguyen2021}. 

While extensive numerical calculations suggest that genuine parton modes do not exist in states belonging to the primary Jain sequence (see Refs.~\cite{Liu18, Liou19} and Appendix~\ref{app: missing_parton_mode}), we demonstrate their existence in the secondary Jain sequence $n/(4n{\pm}1)$ with $n{\geq}2$. For Jain states where CFs have a vorticity $2p{>}4$, we again anticipate the existence of only two spin-$2$ modes, since they too are composed of only two kinds of partons. Thus, we expect that the physics of the $n/(6n\pm1)$ states is similar to that of the $n/(4n{\pm }1)$ states. Furthermore, the bosonic Jain states at $n/[(2p{-}1)n{\pm}1]$ exhibit similar physics, since they are related to the Jain states at $n/(2pn{\pm} 1)$ by a Jastrow factor (for concrete examples, see Appendix~\ref{app: bosons}). In this work, our discussion is restricted to the Jain states, and, in the future, it would be worth exploring the nature of high-energy collective modes of other FQH fluids, in particular, non-Abelian FQH fluids.

It is worth emphasizing that the secondary graviton should be distinguished from other high-energy collective modes, such as those obtained by exciting CFs across multiple CF-Landau-like levels~\cite{Majumder09, Rhone11, Balram13, Majumder14} or excitations in which the CF vorticity is altered~\cite{Peterson04}. The latter high-energy modes typically exist in all Jain fractions, including the Laughlin ones, unlike the modes we identify here, which occur only at Jain fractions $n/(2pn{\pm}1)$ when $n,p{\geq}2$. Moreover, these modes are also characterized by higher spin in the effective theory description \cite{Golkar16, nguyen2018fractional}, which makes them fundamentally different from the modes discussed here. For example, these modes would have a weak response to mass anisotropy quenches considered here, since those primarily excite the quadrupolar spin-2 modes~\cite{Liu18}. 

It would be interesting to further explore the utility of the effective field theory [Eq.~\eqref{eq_parton_theory}] at describing the gapless composite Fermi liquid states at filling $\nu {=}1/(2p)$ as well as their pairing instabilities that would describe gapped states at these even-denominator fillings~\cite{ma2019identification}. When it comes to gapped states in the vicinity of $\nu {=} 1/(2p)$, it would be interesting to perform the wave number expansion to one more order and compute the projected static structure factor to the sixth order in momentum expansion. At this order, the leading large $n$ contribution is of the form $n/p$ \cite{Nguyen17}, and it \emph{cannot} be written as a sum of contributions from the boson and fermion sectors. This calculation requires careful treatment of the framing anomaly generated by integrating out partons as well as Chern-Simons terms \cite{Gromov14a, Gromov15}.

\emph{Note added:}
Recently, we became aware of the work of Nguyen~\emph{et al.}~\cite{Nguyen21a}. The authors of Ref.~\cite{Nguyen21a} also discuss the additional graviton mode in the Jain states at $\nu=2/7$ and $2/9$. 

\section{ Acknowledgments} 
We thank Dung Nguyen for useful comments, Ed Rezayi for generously sharing unpublished numerical data with us, and Nicolas Regnault for assistance with the DiagHam library. We thank the authors of Ref.~\cite{Nguyen21a} for sharing their manuscript with us prior to publication.  A. C. B. acknowledges the Science and Engineering Research Board (SERB) of the Department of Science and Technology (DST) for financial support through the Start-up Grant No. SRG/2020/000154. Z. L. is supported by the National Natural Science Foundation of China through Grant No. 11974014 and by the National Key Research and Development Program of China through Grant No. 2020YFA0309200. A. C. B. and Z. P. thank the Royal Society International Exchanges Award No. IES$\backslash$R2$\backslash$202052 for funding support. Some of the numerical calculations reported in this work were carried out on the Nandadevi supercomputer, which is maintained and supported by the Institute of Mathematical Science’s High-Performance Computing Center. Z. P. acknowledges support by the Leverhulme Trust Research Leadership Award No. RL-2019-015 and by EPSRC Grant No. EP/R020612/1. Statement of compliance with EPSRC policy framework on research data: This publication is theoretical work that does not require supporting research data. A. G. was supported by NSF CAREER Grant No. DMR-2045181.



\appendix 

\section{$\nu{=}2/9$ Jain state}
\label{app:2_9_Jain}

In the main text, we numerically demonstrate the existence of two types of collective modes in the $\nu{=}2/7$ Jain state. Here, we show that similar results are obtained at other fractions belonging to the $\nu{=}n/(4n\pm1)$ series, focusing, in particular, on $\nu{=}2/9$. On the spherical geometry, the $\nu{=}2/9$ Jain state occurs at shift $\mathcal{S}{=}6$. 

\begin{figure}[htb]
  \centering
  \includegraphics[width=0.48\linewidth]{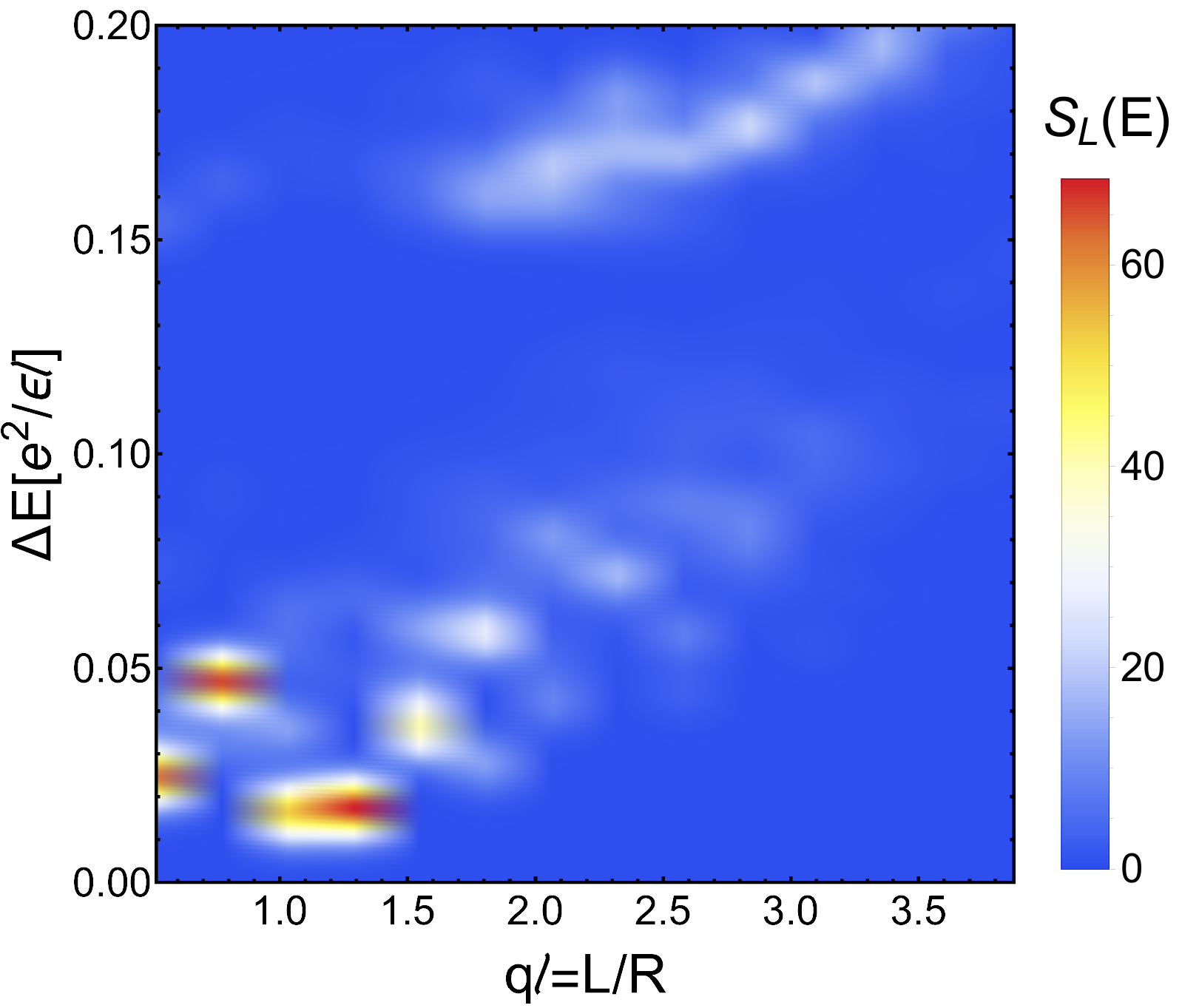}      \includegraphics[width=0.48\linewidth]{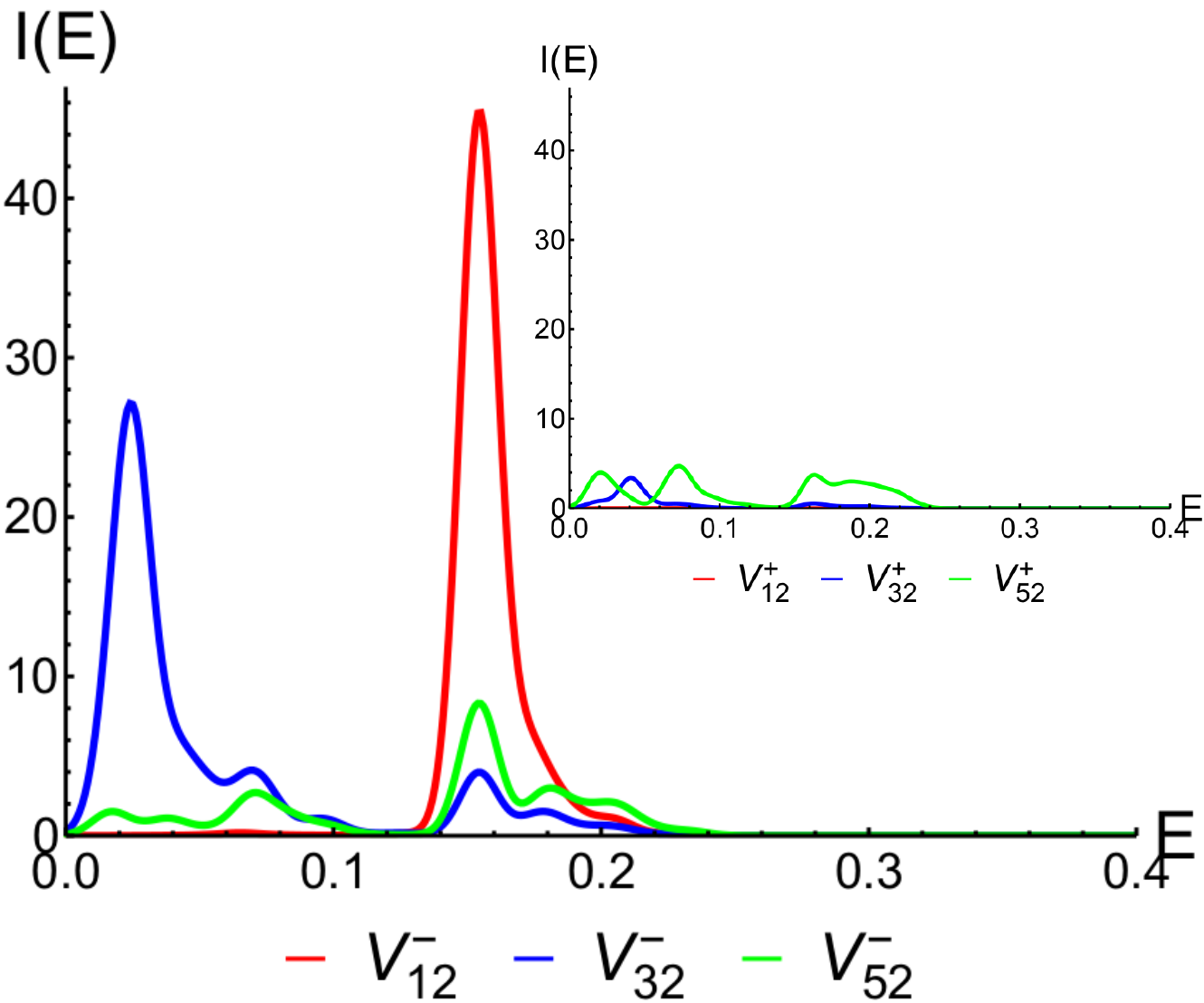}
  \caption{Left: dynamical structure factor $S_L(E)$ at $\nu{=}2/9$ for $N{=}8$ electrons on the sphere, plotted as a function of linear momentum $q\ell{=}L/R$, where $L$ is the angular momentum and $R$ is the sphere radius, and energy $E$ is measured relative to the ground state energy. 
  Right: spectral functions for the generalized pseudopotentials $V_{m,2}^\sigma$ with chirality $\sigma{=}-$ for $\nu{=}2/9$ state. Two dominant peaks, at energies $E{\approx}0.15$ and $E{\approx}0.04$, have the \emph{same} chirality. The inset shows that the analogous spectral functions in $\sigma{=}+$ sectors are strongly suppressed. Data are for $N{=8}$ electrons on the sphere.
  }
  \label{fig:sq29}
\end{figure}
Figure~\ref{fig:sq29} shows the dynamical structure factor and anisotropic response functions evaluated for the $\nu{=}2/9$ state with $N{=}8$ electrons on the sphere. The dynamical structure factor has a large weight on eigenstates near the energy $E{=}0.04$, which corresponds to the GMP collective mode in the long-wavelength limit. In contrast, at higher energies, we do not see clear evidence of the second (parton) mode. However, the parton mode appears clearly in the anisotropic spectral function, shown on the right in Fig.~\ref{fig:sq29}. Here, we observe two pronounced peaks, with energies $E{\approx}0.04$ and $E{\approx}0.15$, in the negative chirality sector. The response in the positive chirality sector, on the other hand, is strongly suppressed -- see the inset to the right in Fig.~\ref{fig:sq29}. The identification of the two modes and their energies is in agreement with the wave functions of the CF-exciton and parton modes shown in Fig.~\ref{fig: CFE_parton_2_9} below. Furthermore, as explained in Sec.~\ref{sec: parton_wf}, the chirality of the parton mode at $\nu{=}2/9$ is indeed expected to be the same as that of the GMP mode, since the effective magnetic field seen by all of the partons is in the same direction as for the electrons.

\begin{figure}[htb]
		\includegraphics[width=\linewidth]{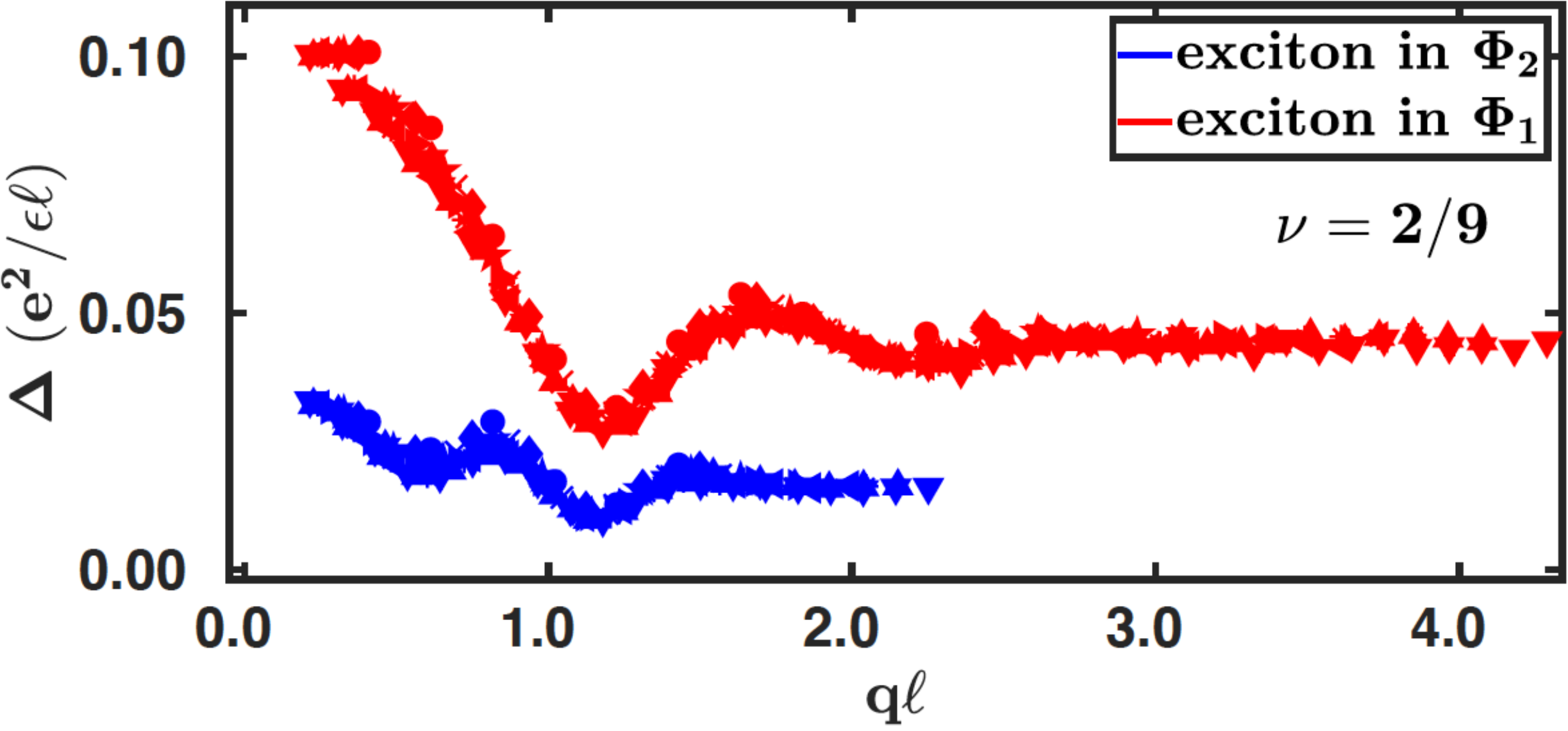}
		\caption{Coulomb energies of the CF exciton and parton modes at $\nu{=}2/9$. Different system sizes are plotted with different symbols with the smallest system with $N{=}12$ electrons and the largest with $N{=}40$. In the long-wavelength limit, the high-energy (red) mode extrapolates to an energy of approximately 0.11, while the low-energy (blue) mode extrapolates to approximately 0.04 in Coulomb units. }
		\label{fig: CFE_parton_2_9}
\end{figure}

\begin{figure}[htb]
		\includegraphics[width=1\linewidth]{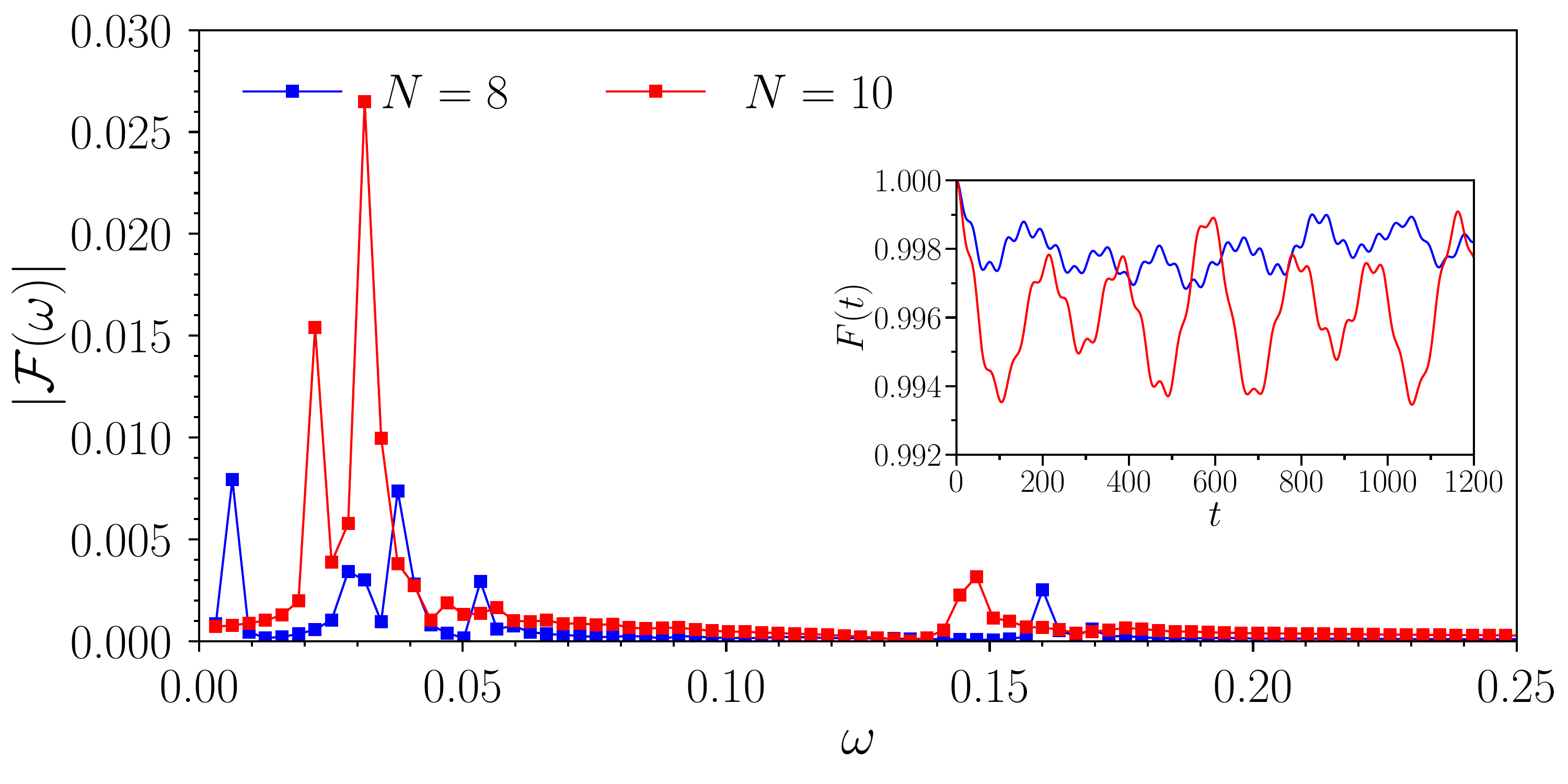}
		\caption{Fourier transform of the postquench fidelity (inset) for the mass anisotropy quench on a torus with the square unit cell. Data are presented for $N{=}8$ and 10 fermions at $\nu{=}2/9$. We choose the Coulomb interaction with $\alpha{=}1.1$.}
		\label{fig:dynamics_2_9}
\end{figure}

The results of the geometric quench for electrons at $\nu{=}2/9$ interacting via LLL-projected Coulomb interaction are shown in Fig.~\ref{fig:dynamics_2_9}. We indeed observe two peaks near the frequencies $\omega{=}0.04$ and $\omega{=}0.15$ in the Fourier transform of the postquench fidelity, supporting the existence of two spin-$2$ modes. While the quench in Fig.~\ref{fig:dynamics_2_9} is simulated using torus geometry, the values of two dominant frequencies are found to be in good agreement with the peaks of the anisotropic spectral functions on the sphere shown in Fig.~\ref{fig:sq29}. Nevertheless, comparing with the sphere spectral functions in Fig.~\ref{fig:sq29}, we also notice that the low-frequency peaks have a considerably larger magnitude than the high-frequency peak. This may be a consequence of small systems that we can simulate using exact diagonalization and the fact that finite-size effects are stronger at $\nu{=}2/9$ compared to $\nu{=}2/7$. Further evidence of the importance of finite-size effects comes from the big difference in the low-frequency peaks between $N{=}8$ and $N{=}10$ electrons, suggesting that larger sizes are necessary to suppress the finite-size fluctuations.

\section{Absence of the parton mode in the $\nu{=}1/5$ Laughlin state and $\nu{=}2/5$ Jain state}
\label{app: missing_parton_mode}
\begin{figure}[bht]
  \centering
  \includegraphics[width=\linewidth]{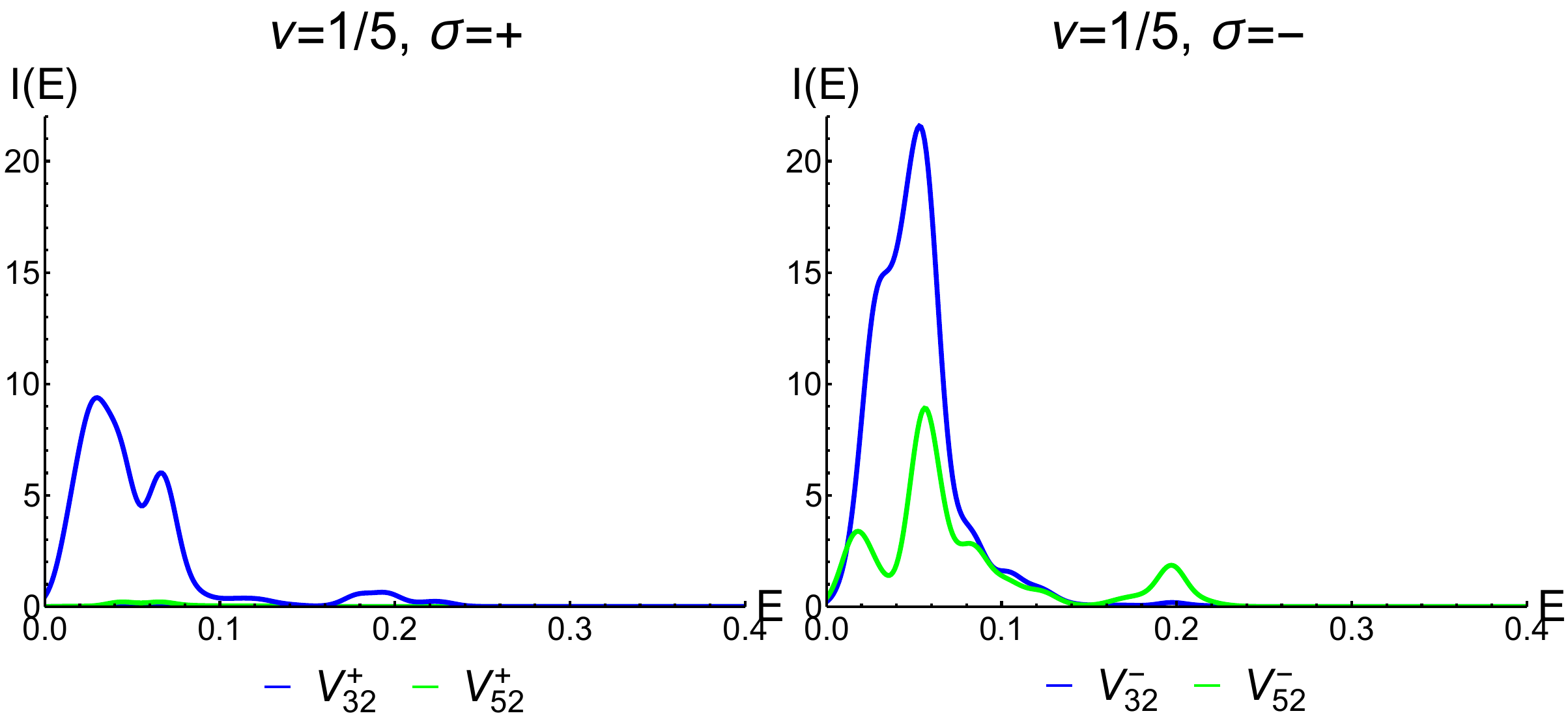}
    \includegraphics[width=\linewidth]{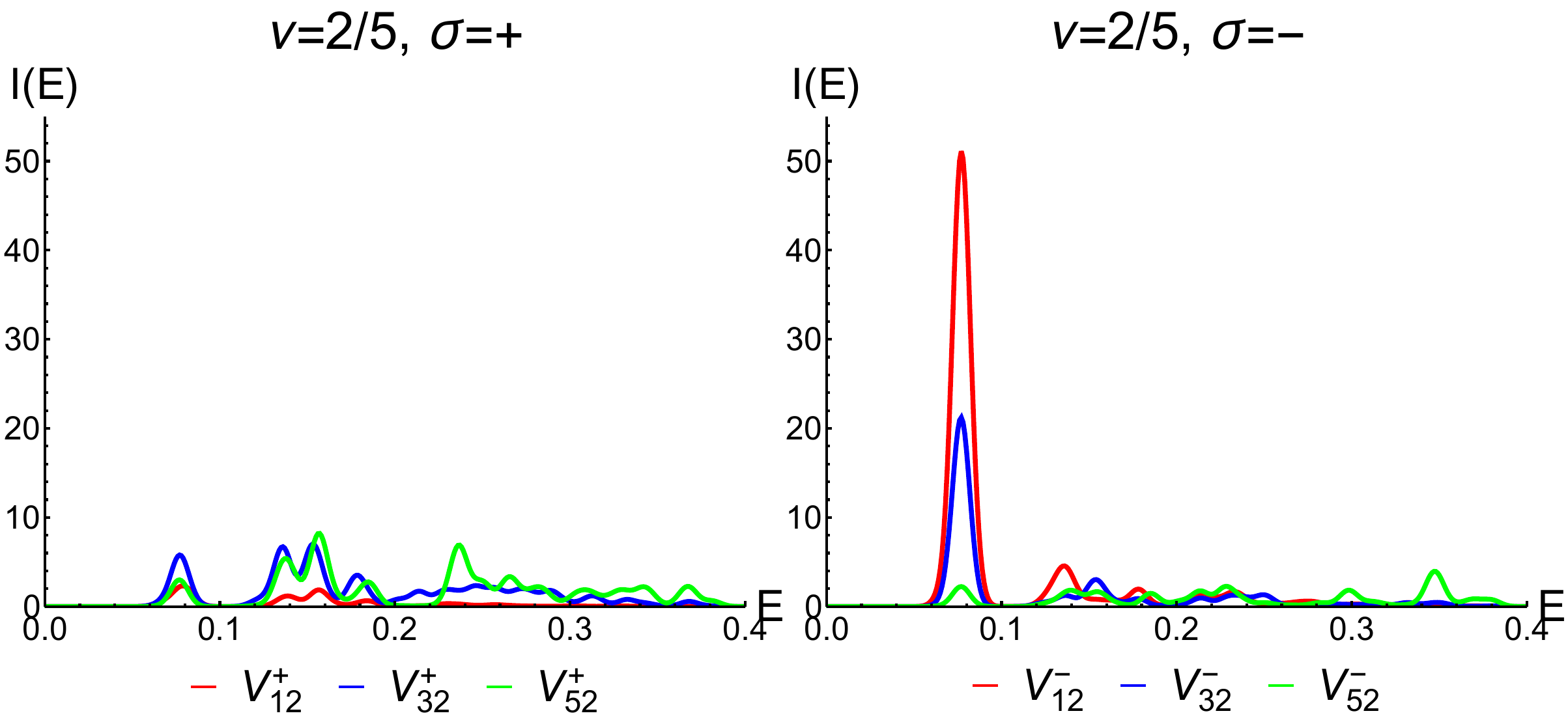}
  \caption{Spectral functions for the anisotropic pseudopotentials $V_{m,2}^\sigma$ with chirality $\sigma$. (a), (b) The $\nu{=}1/5$ Laughlin state has a dominant peak at energy $E{\approx} 0.06$ with \emph{negative} chirality. The peak is broader than in other cases, and the spectral function has about 25\% weight in the positive chirality sector, possibly due to strong finite-size effects. 
  (c), (d) The $\nu{=}2/5$ state has a single dominant peak at energy $E{\approx} 0.08$ with the same chirality at the $\nu{=}1/5$ state. Data are for $N{=7}$ electrons ($\nu{=}1/5)$ and $N{=}10$ electrons ($\nu{=}2/5$) on the sphere. }
  \label{fig:vmn_nu_1_3_nu_2_5}
\end{figure}

The absence of the secondary collective mode is established numerically for the $\nu{=}1/3$ Laughlin state~\cite{Zhao18, Liou19}. Here, we provide similar numerical evidence for the absence of the parton mode in the $\nu{=}1/5$ Laughlin state and $\nu{=}2/5$ Jain state. In Fig.~\ref{fig:vmn_nu_1_3_nu_2_5}, we compute the anisotropic spectral functions for these two states. For both states, the data are consistent with a single dominant peak in the negative chirality sector, the same as that of the $\nu{=}1/3$ state. In the case of the $\nu{=}1/5$ state, we find that the integrated $V_{1,2}$ spectral function (before normalization) is smaller by an order of magnitude compared to those of $V_{3,2}$ and $V_{5,2}$ spectral functions; hence, we omit it from Fig.~\ref{fig:vmn_nu_1_3_nu_2_5}. 

\begin{figure}[htb]
		\includegraphics[width=1\linewidth]{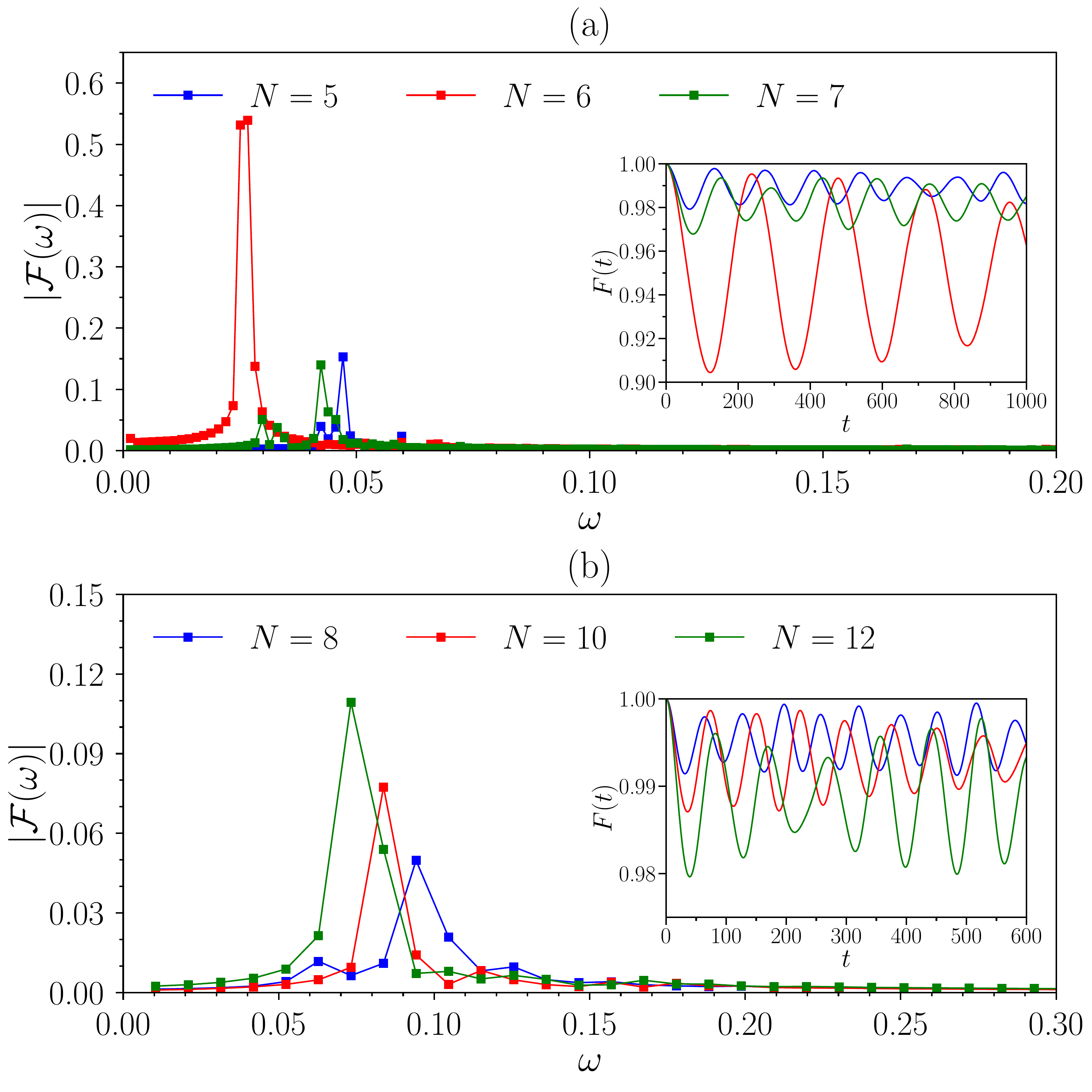}
		\caption{Fourier transform of the postquench fidelity (inset) for the mass anisotropy quench on a torus with the square unit cell. Data are presented for (a) $N{=}5,6,7$ fermions at $\nu{=}1/5$ and (b) $N{=}8,10,12$ fermions at $\nu{=}2/5$. We choose the Coulomb interaction with mass anisotropy $\alpha{=}1.3$ in (a) and $\alpha{=}1.1$ in (b).
		For the $\nu{=}1/5$ state in (a), the different behavior for $N{=}6$ likely results from the electrons' tendency to form a hexagonal Wigner crystal. 	}
		\label{fig:dynamics_1_5_2_5}
\end{figure}
Beyond spectral functions, we look for the existence of the parton mode directly by measuring the response of the FQH system to the geometric quench. At $\nu{=}1/3$, it is shown that such dynamics are governed only by a single frequency corresponding to the energy of the GMP mode in the long-wavelength limit~\cite{Zhao18}. Here, we study the geometric quench also for $\nu{=}1/5$ and $\nu{=}2/5$ FQH states. In Fig.~\ref{fig:dynamics_1_5_2_5}, we can see that the quench dynamics has only one dominant frequency in both cases, thus confirming the absence of the secondary spin-$2$ mode at these fillings. Our numerical results at $\nu{=}1/5$ are consistent with those of Ref.~\cite{Yuzhu21}, where the authors show microscopically that the $1/5$ Laughlin phase has only a single graviton mode.

From the perspective of wave functions, in principle, one can construct states like $\mathcal{P}_{\rm LLL}\Phi_{\pm 2}\Phi_{1}\Phi^{\rm ex}_{1}$. Because of technical difficulties with projection to the LLL, we are not been able to construct these wave functions explicitly. Based on previous numerical calculations for the model Hamiltonian of the 1/3 Laughlin state as well as the Coulomb interaction (where there is evidence for only a single mode~\cite{Zhao18}), which apply to 2/3 by particle-hole conjugation, we conjecture that the aforementioned state is projected out. If the state $\mathcal{P}_{\rm LLL}[\Phi_{2}]^{*}\Phi_{1}\Phi^{\rm ex}_{1}$ is a valid spin-$2$ state, then there will be two spin-$2$ modes at 2/3 (the CFE mode is seen in the spectral function), which is not consistent with the previous numerical findings. This premise is also consistent with the effective field theory, where it is shown that a Dirac fermion at $n/(2n{\pm}1)$, for large $n$, hosts only a single collective mode with spin-$2$~\cite{Golkar16}.

Finally, we also provide a heuristic argument for the absence of the parton modes. Among the entire family of parton states (here, we can also include the composite fermion states) that have been considered to date, the ones that seem to be relevant are the ones that can be projected into the LLL using the JK method~\cite{Jain97b}. The wave function $\mathcal{P}_{\rm LLL}\Phi_{\pm 2}\Phi_{1}\Phi^{\rm ex}_{1}$, due to a paucity of the Jastrow factors, cannot be projected into the LLL state using the JK method. On the other hand, for the states that we construct, projection into the LLL can be carried out using the JK method as shown in Eq.~\eqref{eq: parton_mode}. Essentially, enough factors of $\Phi_{1}$ that facilitate a projection into the LLL using the JK method likely help in building good correlations in the wave function~\cite{Balram18a}.

\section{Bosons}
\label{app: bosons}

While FQH states are naturally formed by electrons, they can also arise in systems of bosons, e.g., realized by ultracold atoms -- see, e.g., Refs.~\cite{CooperWilkinGunn, RegnaultJolicoeur} and the recent review in Ref.~\cite{HalperinJainBook}. For any fermionic FQH state, an analogous bosonic FQH state can be obtained by dividing its wave function through with an overall Jastrow factor $\Phi_{1}$, which leads to a fully symmetric wave function under exchange of any two particles. The filling factors of the corresponding fermionic and bosonic FQH states are related by $\nu_f^{-1} {=} \nu_b^{-1} + 1$. In particular, the bosonic Jain state at $\nu{=}n/[(2p-1)n \pm 1]$ is related to its fermionic counterpart at $\nu{=}n/(2pn \pm 1)$. Here, we focus on the bosonic states $\nu{=}2/5$ and $\nu{=}2/7$, whose fermionic analogs are $\nu{=}2/7$ and $\nu{=}2/9$, respectively. The Wen-Zee shift of these bosonic states can similarly be shown to be $\mathcal{S}{=}1$ and $\mathcal{S}{=}5$, respectively. For general sequences of Jain states, the physical properties of bosonic versions of Jain states are found to largely mirror those of fermionic Jain states~\cite{CooperWilkin, RegnaultJolicoeur}. 

\begin{figure}[htb]
  \centering
  \includegraphics[width=0.48\linewidth]{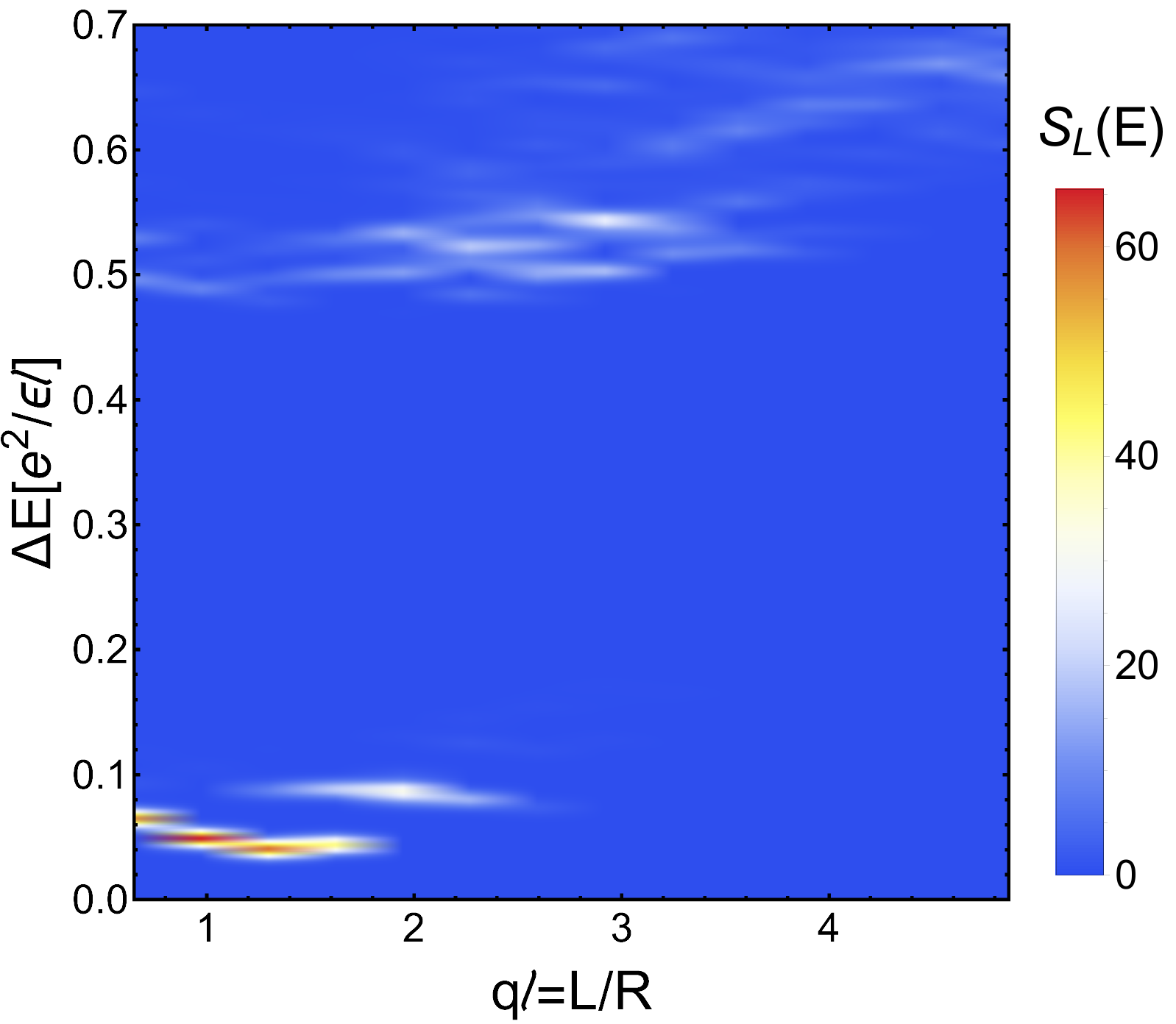}
    \includegraphics[width=0.48\linewidth]{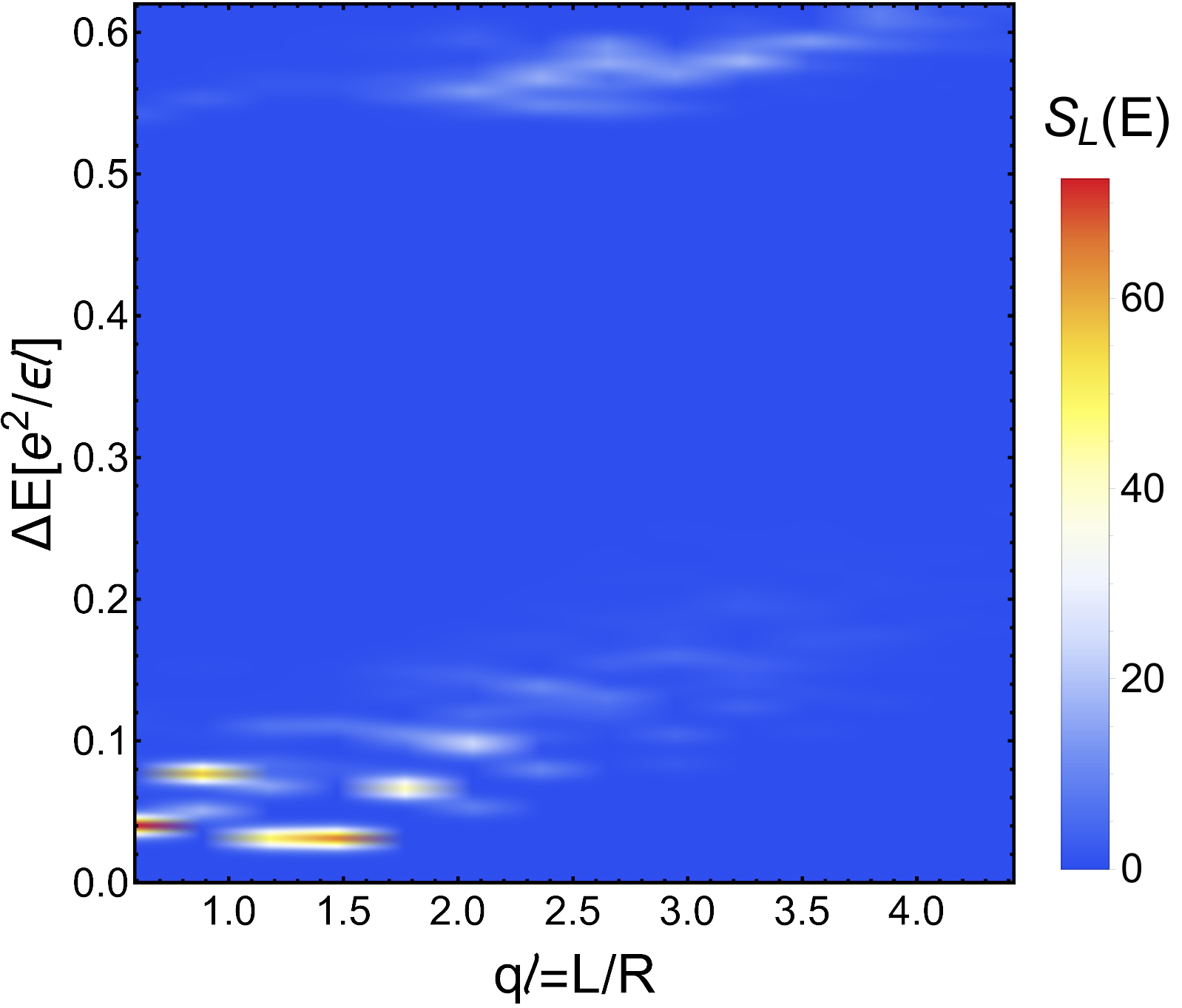}
  \caption{Dynamical structure factor $S_L(E)$ on the sphere geometry plotted as a function of linear momentum $q\ell{=}L/R$, where $L$ is the angular momentum and $R$ is the sphere radius, and energy $E$ measured relative to the ground state, in units of $e^2/(\epsilon\ell)$. Left: $N{=}8$ bosons at filling $\nu{=}2/5$, corresponding to fermionic filling $\nu{=}2/7$. Right: $N{=}8$ bosons at filling $\nu{=}2/7$, corresponding to fermionic filling $\nu{=}2/9$. }
  \label{fig:sq_bosons}
\end{figure}
In Fig.~\ref{fig:sq_bosons}, we show the dynamical structure for bosons at $\nu{=}2/5$ and $\nu{=}2/7$ on the spherical geometry. In both cases, most of the spectral weight is carried by the GMP mode at low energies ($E{\lesssim}0.1$). In the case of $\nu{=}2/5$, we also observe an additional mode in the long-wavelength limit at energy $E{\approx}0.5$. On the other hand, at $\nu{=}2/7$ we observe only weak signatures of the second mode, similar to the fermionic case at $\nu{=}2/9$ in Fig.~\ref{fig:sq29}. 

\begin{figure}[b]
  \centering
  \includegraphics[width=\linewidth]{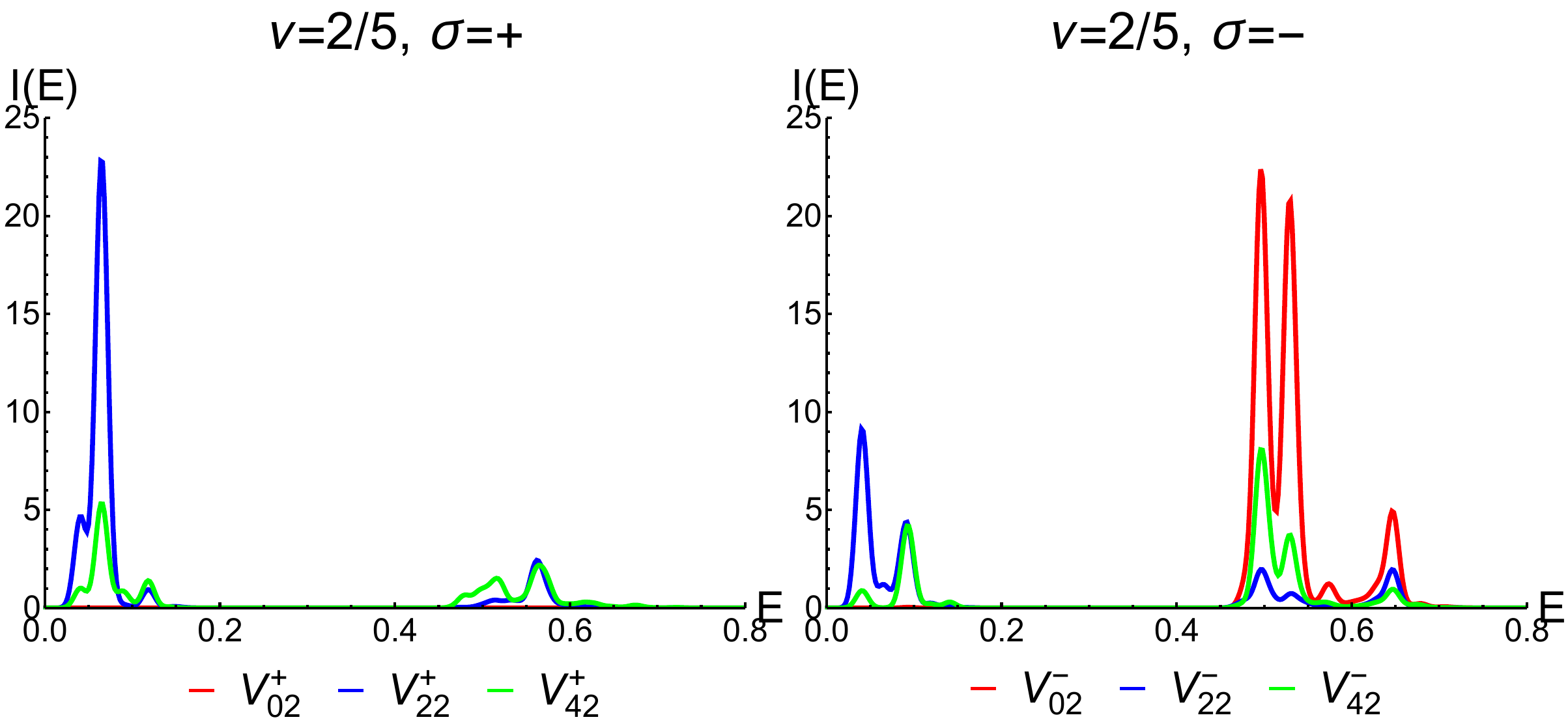}
  \includegraphics[width=\linewidth]{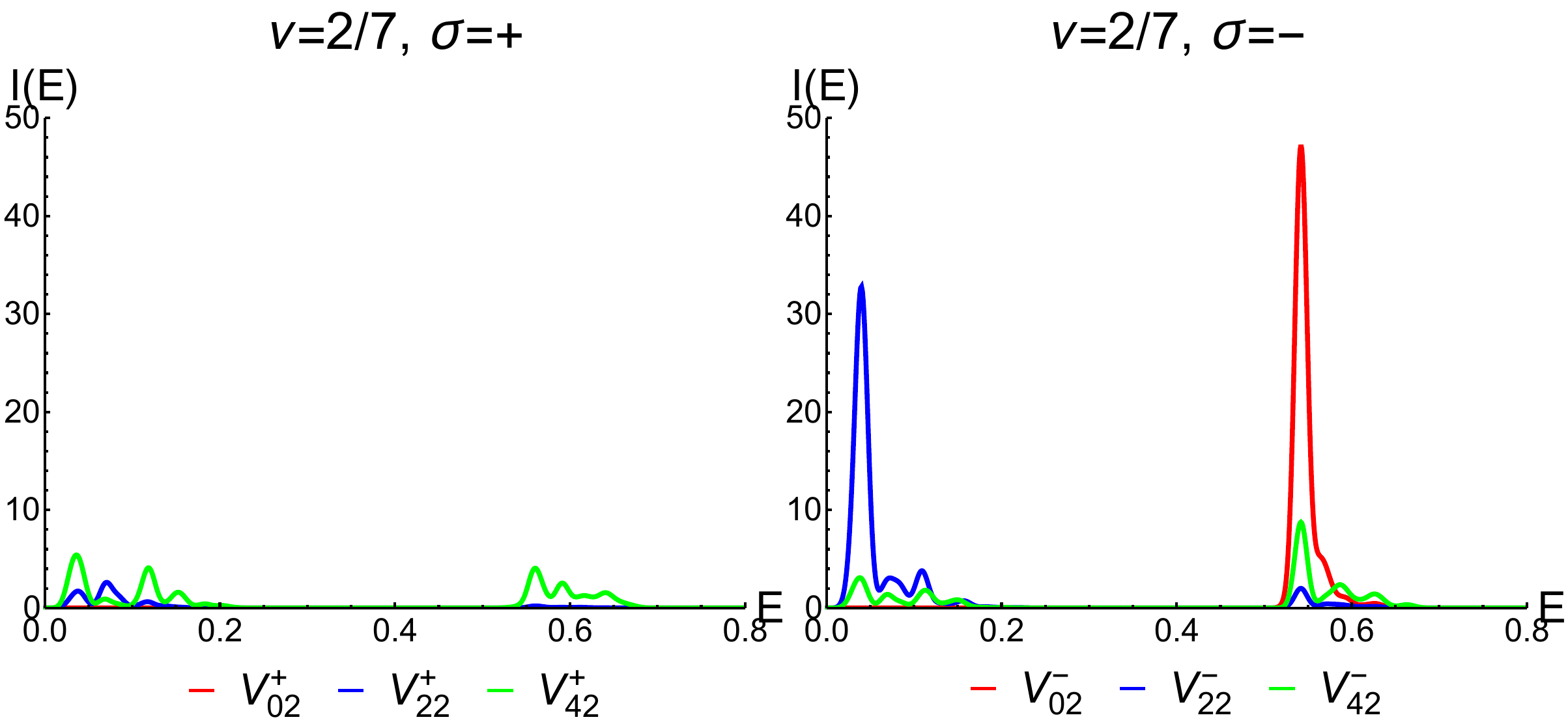}
  \caption{Spectral functions for the generalized pseudopotentials $V_{m,2}^\sigma$ with chirality $\sigma$ for bosons at $\nu{=}2/5$ and $\nu{=}2/7$, corresponding, respectively, to fermionic filling factors $\nu{=}2/7$ and $\nu{=}2/9$. Similar to the fermionic case, we observe two dominant peaks at both filling factors, corresponding to two collective modes. The chirality of the two modes is different (the same) at $\nu{=}2/5$ ($\nu{=}2/7$). Data is for $N{=8}$ electrons on the sphere. }
  \label{fig:vmnbosons}
\end{figure}

Clearer signatures of the two modes can be seen in the anisotropic spectral functions shown in Fig.~\ref{fig:vmnbosons}. Recall that these spectral functions are evaluated from the matrix element of anisotropic pseudopotentials, $V_{m,s}^\sigma$, where we fix $s{=}2$ to probe quadrupolar response and $m{=}0,2,4,\ldots$ is even, since the wave function for bosons must be symmetric. At both filling fractions, we observe two pronounced peaks in the spectral response, occurring roughly at energies $E{\approx}0.05$ and $E{\approx}0.5$. The two modes corresponding to these peaks have opposite chirality at $\nu{=}2/5$ and the same chirality at $\nu{=}2/7$, similar to the corresponding results for fermions. We note that the variational wave functions, based on CF excitons and partons, can be directly generalized to the bosonic case by dividing the fermionic wave functions by a factor of $\Phi_1$. The variational energy of such wave functions is found to match closely the energies at which the peaks occur in Fig.~\ref{fig:vmnbosons} (data not shown).

\begin{figure}[htb]
		\includegraphics[width=\linewidth]{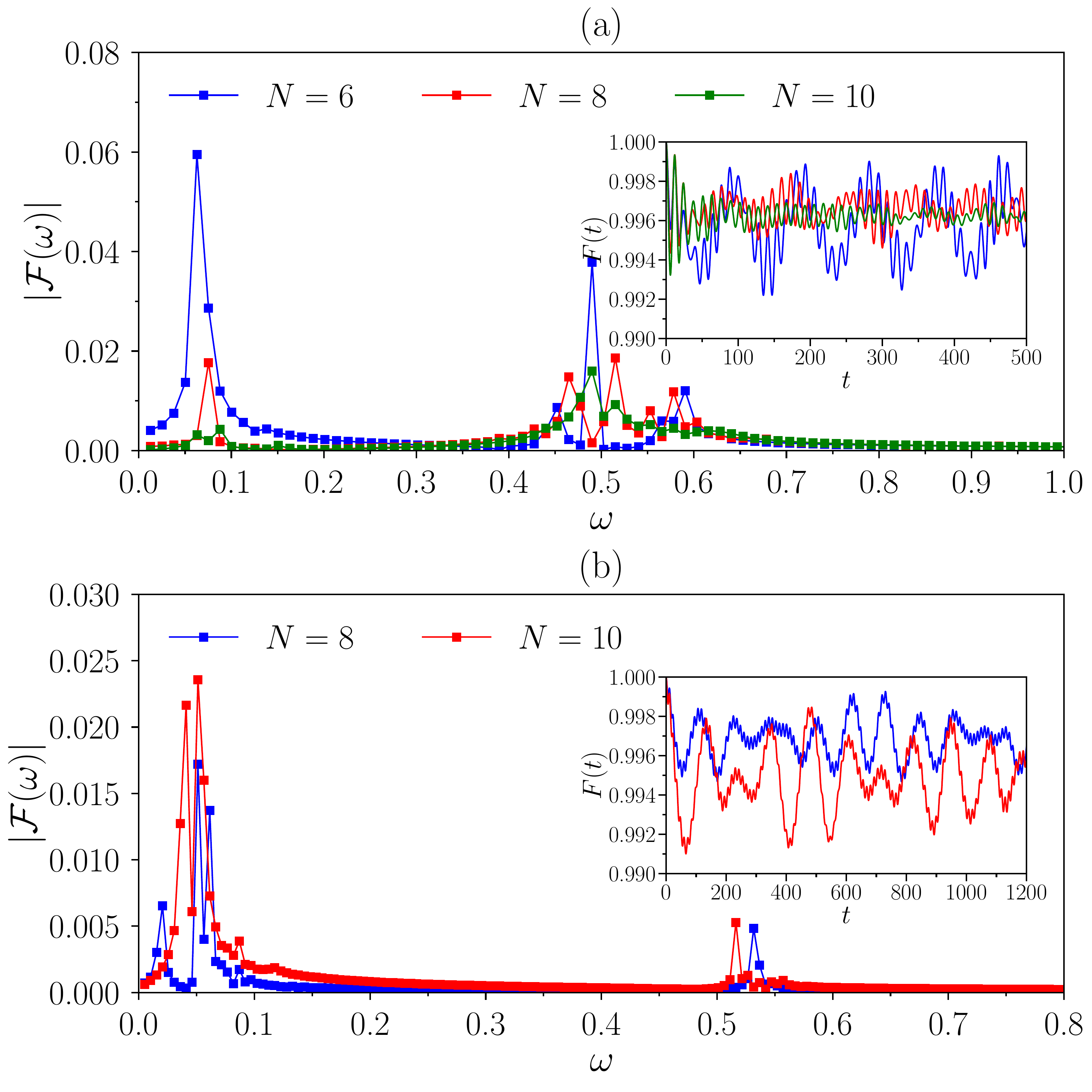}
		\caption{Fourier transform of the postquench fidelity (inset) for the mass anisotropy quench on a torus with the square unit cell. Data are for (a) $N{=}6,8,10$ bosons at $\nu{=}2/5$ and (b) $N{=}8,10$ bosons at $\nu{=}2/7$. We choose Coulomb interaction with mass anisotropy $\alpha{=}1.1$.}
		\label{fig:dynamics_bosons}
\end{figure}

Finally, we also study geometric quenches for bosonic Jain states at $\nu{=}2/5$ and $\nu{=}2/7$. In the Fourier transform of the postquench fidelity of bosons interacting via Coulomb potential (Fig.~\ref{fig:dynamics_bosons}), we observe clear signatures of two dominant frequencies at both filling factors. These frequencies further match the long-wavelength limits of the two collective modes extracted from the parton and CF-exciton wave functions and are also in agreement with the dynamical structure factor, thus confirming the existence of two spin-$2$ degrees of freedom, similar to the fermionic case in Figs.~\ref{fig:dynamics_2_7} and \ref{fig:dynamics_2_9}.  

\section{Calculation of the projected static structure factor}
\label{app: EFT}

To compute the static structure factor, we use the large $n$ approach developed in Refs.~\cite{Nguyen17b, golkar2016higher, Nguyen2021}. Namely, we turn off the background geometry and treat the Dirac parton, in the large $n$ approximation, using collective variables $u_{\pm l}$ describing the angular momentum $l$ distortion of the Dirac parton Fermi surface. We outline the calculation, referring the reader to the original works for a detailed discussion of the Dirac CF Fermi liquid theory. Below, we show that the calculation of the static structure factor completely factorizes and there are just two independent contributions: one coming from the Fermi liquid, and the other coming from the bimetric theory. 

To implement this program, we need the following ingredients from Ref.~\cite{Nguyen2021}. First, the equations of motion for the collective variables are

\begin{align}
&\dot{u}_0 = - v_F\left(\p u_1 - \bar \p u_{-1}\right)\,,
\\
&\dot{u}_{1} = -i(1+F_1) \frac{\bar b v_F}{k_F} u_1 - v_F(1+F_0) \p u_0 
\\ \nonumber
&- v_F(1+F_2) \bar \p u_2 + \bar e,
\\
&\dot{u}_2 = -i(1+F_2)\frac{\bar b v_F}{k_F} u_2\,,\qquad u_{-i} = u_i^\star\,,
\end{align}
where $F_i$ are the Landau parameters and $v_F$ and $k_F$ are the Fermi velocity and momentum, respectively. The latter is determined by the Luttinger theorem \cite{Nguyen2021}, which is expected to hold for composite partons \cite{wang2019dirac}
\be
\frac{k_F^2}{4\pi} = \bar \rho_{CF} = \frac{1}{4\pi} \frac{p-1}{p} \bar b + \frac{B}{4\pi p}\,,
\ee
where $\hat R$ is the curvature of the metric $\hat{g}_{ij}$.
We also need the relation between composite fermion density and $u_0$:
\be
\rho_{CF} = \frac{k_F}{2\pi} u_0 \Rightarrow \dot{u}_0 = \frac{p-1}{2k_Fp} \dot{b} + \frac{\varsigma}{2p} \frac{\p_0 \hat R}{2k_F}\,.
\ee
We are interested in expressing the fluctuation of electron density $\delta \rho_e$ in terms of variables $u_{{\pm} l}$ and $\hat R$. To do this, first observe that the electron density is given by
\be
\rho_e = \frac{B-b}{4\pi p} + \frac{i k_F^2}{2\pi B}\left(\p u_1 - \bar \p u_{-1}\right) + \frac{1}{4\pi}\frac{\varsigma}{2p} \hat R\,.
\ee
Following the steps from Ref.~\cite{Nguyen2021} we get a simple expression for the time derivative of $\rho_e$:
\be
\dot{\rho}_e = - i \frac{v_Fk_F^2}{2\pi B} (1+F_2)(\p \dot{u_2} - \bar \p \dot{u}_{-2}) + \frac{\varsigma}{2p} \frac{\p_0 \hat R}{4\pi} \left(1 - \frac{\bar b}{B}\right)\,.
\ee
Finally, removing the time derivative and using that $\bar b {=} B/(2pn{+}1)$, we find that the density fluctuation is given by
\be\la{eq_rho}
\delta \rho_e = - i \frac{v_Fk_F^2}{2\pi B} (1+F_2)(\p u_2 - \bar \p u_{-2}) + \frac{\nu \varsigma}{4\pi} \hat R.
\ee
The calculation of the static structure factor
\be
\bar{s}(q) = \frac{1}{\bar{\rho}_e}\langle \delta \rho_e(-q) \delta \rho_e(q) \rangle
\ee
completely factorizes. The first term is studied in Ref.~\cite{Nguyen2021}, whereas the second term is exactly the same as in the bimetric theory \cite{Gromov17}, which results in
\be
\bar s_4 = \frac{1}{8} \left( n+1 + \varsigma \right) = \frac{1}{8}\left( n+2p -1\right)\,.
\ee
The last equation is the known value of $\bar{s}_4$. This fixes the value of $\varsigma$:
\be
\varsigma = p-1 \Rightarrow s=0\,.
\ee
Thus, at least to the lowest order in gradients, the two theories do not interact with each other. Using SMA on the Dirac Fermi liquid theory, we can decouple the spin-$2$ mode from the rest of the collective modes $u_{{\pm} l}$ with $l{>}2$. In this limit \cite{Nguyen17b}, Dirac CF Fermi liquid theory is equivalent to a bimetric theory with $\varsigma^\prime {=} (n{+}1)/8$. As a result, we expect that the geometric quench dynamics, studied in Sec.~\ref{sec: quench_dynamics}, is going to be a sum of two oscillations with frequencies corresponding to the gaps of the two spin-$2$ modes.

\bibliography{biblio_fqhe}
\end{document}